\documentclass[a4paper,12pt]{article}
\pdfoutput=1
\usepackage{amsmath,amssymb,amsfonts,cite,color,epsf,epsfig,epstopdf,graphics,graphicx,mathtools,subcaption,physics,verbatim}
\def\thefootnote{*\arabic{footnote}}
\definecolor{ultramarine}{rgb}{0.07, 0.04, 0.56}
\definecolor{cadmiumgreen}{rgb}{0.0, 0.42, 0.24}
\definecolor{indigo(dye)}{rgb}{0.0, 0.25, 0.42}
\usepackage[linktocpage=true,breaklinks]{hyperref}
\hypersetup{
	colorlinks=true,
	citecolor=ultramarine,
	linkcolor=cadmiumgreen,
	urlcolor=indigo(dye),
}

\numberwithin{equation}{section}

\usepackage{tikz}
\usepackage{ragged2e}
\usepackage{fontenc}
\usepackage{listings}
\usepackage{hyperref}
\usepackage{geometry}

\begin{document}

\begin{flushright} {\footnotesize YITP-20-124, IPMU20-0104}  \end{flushright}
\vspace{5mm}
\begin{center}

\def\thefootnote{\fnsymbol{footnote}}

{\Large {\bf Inflationary gravitational waves in consistent $D\to 4$ Einstein-Gauss-Bonnet gravity}}
\\[1cm]

{Katsuki Aoki$^{1}\footnote{katsuki.aoki@yukawa.kyoto-u.ac.jp}$, Mohammad Ali Gorji$^{1}\footnote{gorji@yukawa.kyoto-u.ac.jp}$, 
Shuntaro Mizuno$^{2}\footnote{mizuno-g@hachinohe.kosen-ac.jp}$,
Shinji Mukohyama$^{1,3}\footnote{shinji.mukohyama@yukawa.kyoto-u.ac.jp}$}
\\[.7cm]

{\small \textit{$^1$Center for Gravitational Physics, Yukawa Institute for Theoretical
		Physics, Kyoto University, \\
		606-8502, Kyoto, Japan
}}\\

{\small \textit{$^2$ Department of Liberal Arts and Engineering Sciences, National Institute of Technology, \\
Hachinohe college, 039-1192, Aomori, Japan
}}\\

{\small \textit{$^3$Kavli Institute for the Physics and Mathematics of the Universe (WPI), The University of Tokyo, 277-8583, Chiba, Japan}}\\

\end{center}

\vspace{.8cm}

\hrule \vspace{0.3cm}

\begin{abstract} 
We study the slow-roll single field inflation in the context of the consistent $D\to4$ Einstein-Gauss-Bonnet gravity that was recently proposed in \cite{Aoki:2020lig}. In addition to the standard attractor regime, we find a new attractor regime which we call the Gauss-Bonnet attractor as the dominant contribution comes from the Gauss-Bonnet term. Around this attractor solution, we find power spectra and spectral tilts for the curvature perturbations and gravitational waves (GWs) and also a model-independent consistency relation among observable quantities. The Gauss-Bonnet term provides a nonlinear $k^4$ term to the GWs dispersion relation which has the same order as the standard linear $k^2$ term at the time of horizon crossing around the Gauss-Bonnet attractor. The Gauss-Bonnet attractor regime thus provides a new scenario for the primordial GWs which can be tested by observations. Finally, we study non-Gaussianity of GWs in this model and estimate the nonlinear parameters $f^{s_1s_2s_3}_{\rm NL,\;sq}$ and $f^{s_1s_2s_3}_{\rm NL,\;eq}$ by fitting the computed GWs bispectra with the local-type and equilateral-type templates respectively at the squeezed limit and at the equilateral shape. For helicities $(+++)$ and $(---)$, $f^{s_1s_2s_3}_{\rm NL,\;sq}$ is larger while
$f^{s_1s_2s_3}_{\rm NL,\;eq}$ is larger for helicities $(++-)$ and $(--+)$.
\end{abstract}
\vspace{0.5cm} 
\hrule
\def\thefootnote{\arabic{footnote}}
\setcounter{footnote}{0}

\newpage

\section{Introduction}\label{introduction}

According to the Lovelock theorem~\cite{Lovelock:1970,Lovelock:1972vz}, under a set of reasonable assumptions, the local gravitational equation of motion for a four dimensional metric is uniquely given by the Einstein equation with or without the cosmological constant. Therefore, in order to modify general relativity (GR), one needs to break at least one of the assumptions in the Lovelock theorem. Nonetheless, it was recently proposed in \cite{Glavan:2019inb} that one could obtain a covariant gravitational equation of motion for a four dimensional metric by taking a singular $D\to 4$ limit of Einstein-Gauss-Bonnet gravity in $D>4$ dimensions, that the theory would propagate only two local physical degrees of freedom and that the theory would be different from GR. As a statement about the equation of motion for a four-dimensional metric, this proposal explicitly contradicts with the Lovelock theorem. Indeed, there have been many papers in the literature that pointed out inconsistencies and ambiguities of the proposal. For example, the resulting equations of motion in four dimensions are not regular in general \cite{Gurses:2020ofy} (see also \cite{Ai:2020peo,Mahapatra:2020rds,Arrechea:2020evj}) and there is no regular action that reproduces the suggested regularized equations of motion \cite{Hohmann:2020cor}. It was also shown that taking the limit of the action from higher dimensions, in the simplest case one ends up with a scalar-tensor theory \cite{Lu:2020iav,Kobayashi:2020wqy,Ma:2020ufk}. It is possible to reproduce the solutions that were found in \cite{Glavan:2019inb} while the extra scalar degree of freedom becomes strongly coupled around those solutions \cite{Kobayashi:2020wqy} and, therefore, one cannot trust the solutions of \cite{Glavan:2019inb} in its scalar-tensor realizations suggested in \cite{Lu:2020iav,Kobayashi:2020wqy,Ma:2020ufk}. The appearance of this extra scalar mode under the limit $D\to4$ is also deduced from different perspectives through the calculations of the gravitational amplitudes in $D>4$ dimensional spacetime \cite{Bonifacio:2020vbk} and also a regularization based on a conformal transformation \cite{Fernandes:2020nbq,Hennigar:2020lsl,Easson:2020mpq}. Moreover, under the general criterion of Lorentz invariance, unitarity and locality, it is shown that there is not any new gravitational amplitudes for the limit $D\to4$. However, the more fundamental issue is that the limit $D\to4$ is not unique in general so that, depending on the properties of the extra dimensional space, many models with many extra degrees of freedom would appear. Therefore, it was not possible to define the theory completely in a four dimensional spacetime without any reference to the information about the extra dimensions. 

Ref.~\cite{Aoki:2020lig}, after taking into account all those issues, proposed a consistent theory of $D\to4$ Einstein-Gauss-Bonnet gravity that propagates only two local physical degrees of freedom at the fully nonlinear level. First, it was confirmed that taking the limit $D\to4$ either breaks the diffeomorphism invariance or leads to extra degree(s) of freedom in agreement with the Lovelock theorem. Second, as we explained above, in the simplest direct realization of $D\to4$ limit with one scalar degree of freedom, the resultant theory is not new if it is not ill-defined. In practice, the $D\to 4$ limit with the extra degree(s) of freedom is generically pathological due to the infinite strong coupling as discussed in~\cite{Aoki:2020iwm} unless a different scaling limit of $\alpha$ is considered~\cite{Bonifacio:2020vbk}. Therefore, the only possibility to realize a new consistent formulation of the $D\to4$ Einstein-Gauss-Bonnet theory with two local physical degrees of freedom is to break the diffeomorphism invariance. Finally, the construction of a four dimensional theory along this line was systematically and consistently done in \cite{Aoki:2020lig} by breaking only time diffeomorphism in the context of the minimally modified gravity models \cite{DeFelice:2015hla,Lin:2017oow,Aoki:2018zcv,FKMM,Aoki:2018brq,Mukohyama:2019unx,DeFelice:2020eju}. 

For the consistent version of $4$-dimensional Einstein-Gauss-Bonnet (4DEGB) gravity, since there is not time diffeomorphism in the theory, it is natural to work with the Arnowitt-Deser-Misner (ADM) formalism with the $4$-dimensional metric of the form
\begin{equation}\label{metric}
ds^2 = g_{\mu\nu} dx^{\mu} dx^{\nu} = - N^2 dt^2 + \gamma_{ij} ( dx^i + N^i dt ) ( dx^j + N^j dt ) \,,
\end{equation}
in which $N$ is the lapse function, $N^i$ is the shift vector, and $\gamma_{ij}$ denotes the spatial metric. All matter fields would then minimally couple to the above metric. The gravitational action is given by \cite{Aoki:2020lig}
\begin{eqnarray}\label{action-EGB}
S_{\rm g} &=& \frac{M_{\rm Pl}^2}{2} \int dt d^3x N\sqrt{\gamma} 
\Big[ 2R - \mathcal{M} + \tilde{\alpha} R^2_{\rm 4DGB} \Big], 
\\
R^2_{\rm 4DGB} &\equiv& \frac{1}{2}\Big( 8R^2 -4 R\mathcal{M} -\mathcal{M}^2
- \frac{8}{3} \big(8R_{ij}R^{ij}-4R_{ij}\mathcal{M}^{ij}
-\mathcal{M}_{ij}\mathcal{M}^{ij}\big) \Big) ,
\end{eqnarray}
where 
\begin{eqnarray}\label{M}
\mathcal{M}_{ij} \equiv R_{ij}+\mathcal{K}^k{}_{k} \mathcal{K}_{ij}-\mathcal{K}_{ik}\mathcal{K}^k{}_{j}, 
\hspace{1cm} \mathcal{M} \equiv \mathcal{M}^i{}_{i} \,, 
\end{eqnarray}
and
\begin{eqnarray}\label{K}
\mathcal{K}_{ij} \equiv \frac{1}{2N}( \dot{\gamma}_{ij}-2D_{(i}N_{j)}-\gamma_{ij}D^2 \lambda_{\rm GF} ) \,.
\end{eqnarray}
Here, a dot denotes the derivative with respect to $t$. The theory defined by the action (\ref{action-EGB}) has the following five properties: (i) it is invariant under 3-dimensional spatial diffeomorphism; (ii) the number of the local physical degrees of freedom is two as in GR; (iii) it reduces to GR for $\tilde{\alpha}=0$; (iv) all Gauss-Bonnet corrections $R^2_{\rm 4DGB}$ are 4th-order in derivatives; (v) if the Weyl tensor of the spatial metric and Weyl piece of $K_{ik}K_{jl}-K_{il}K_{jk}$, where $K_{ij}=\frac{1}{2N}( \dot{\gamma}_{ij}-2D_{(i}N_{j)})$, vanish for a solution of $D$-dimensional Einstein-Gauss-Bonnet gravity, the $D\rightarrow 4$ limit of this solution is a solution of 4DEGB theory defined by the action (\ref{action-EGB}). The last condition makes it quite reasonable to consider this well-defined theory as $4$-dimensional realization of the Einstein-Gauss-Bonnet gravity.

In addition to the spatial diffeomorphism, the action (\ref{action-EGB}) also enjoys the time reparametrization symmetry while it does not respect the full spacetime diffeomorphism and it is Lorentz violating. Clearly, the time diffeomorphism is broken only with the Gauss-Bonnet corrections and, therefore, Lorentz violation is suppressed by $\tilde{\alpha}$ in the gravity sector. The Lorentz violation would also penetrate to the matter sector through the graviton loops which is suppressed not only by $\tilde{\alpha}$ but also by negative powers of $M_{\rm Pl}^2$. Therefore, the Lorentz violation in the theory is under control. The conditions (i)-(v) then uniquely determine the consistent 4DEGB theory with the action (\ref{action-EGB}) up to a choice of the gauge-fixing constraint which is enforced by the Lagrange multiplier $\lambda_{\rm GF}$. The existence of the gauge-fixing term is necessary to have a consistent Lorentz-violating theory \cite{Aoki:2020lig}. We have chosen the gauge-fixing constraint so that it can be compatible with the cosmological solution and the black hole solution suggested by \cite{Glavan:2019inb}. The gauge-fixing constraint used in the present paper reduces to the constant mean curvature slice $K=K(t)$ when we take the GR limit $\tilde{\alpha}\to0$. In the case of homogeneous cosmological backgrounds and also  spherically symmetric backgrounds but with any value of $\tilde{\alpha}$, the constraint is trivially satisfied and one can set $\lambda_{\rm GF}$ to zero in practice and (\ref{K}) reduces to the standard extrinsic curvature. Moreover, in Ref. \cite{Aoki:2020iwm} we showed that also for linear cosmological perturbations $\lambda_{\rm GF}(t,{\bf x})$ does not play any role and one can set it to zero in practice. However, for nonlinear perturbations $\lambda_{\rm GF}(t,{\bf x})$ cannot set to zero in general. 

Having constructed a new modified gravity theory, it is quite interesting to look at its application in cosmology and black hole physics. In the case of cosmology, as a first step, the linear cosmological perturbations were studied in Ref. \cite{Aoki:2020iwm} and it was shown that they are free of any disastrous instabilities and that interestingly the dispersion relation of the tensor modes receives nonlinear corrections from the Gauss-Bonnet term. Therefore, as a next step, in the present paper we shall study linear and also nonlinear primordial gravitational waves (GWs) in this scenario to look for the nontrivial effects of the Gauss-Bonnet term during inflation. Indeed, non-Gaussianity of GWs was recently studied in the context of modified gravity theories \cite{Gao:2011vs,Goon:2018fyu,Fujita:2019tov,Bordin:2020eui,Bartolo:2020gsh}. The spectrum of the GWs is independent of the inflationary model and therefore we will be able to find some model-independent properties of the inflationary scenarios based on the 4DEGB gravity. In this regard, we find the characteristic properties of the theory which makes it different than GR and also other modified theories of gravity.

The rest of the paper is organized as follows: In Section \ref{sec-inflation}, we consider the standard slow-roll inflation in the consistent 4DEGB gravity framework and we obtain the background equations and power spectrum of the curvature perturbations. In Section \ref{sec_tensor} we study linear tensor perturbations and we find the power spectrum for the primordial GWs. In Section \ref{sec-NG}, we study the non-Gaussianity of the primordial GWs. In Section \ref{sec-obs} we find a model-independent consistency relation for the inflationary observables and we discuss the implications of CMB observational bounds. Section \ref{summary} is devoted to the summary and discussions and we present some details of calculations in appendices \ref{app-Whittaker} and \ref{app-polarization}.

\section{Single field inflation}\label{sec-inflation}

In this section, we consider slow-roll single field inflation in the consistent version of 4DEGB gravity. The action for the inflaton scalar field $\phi$ with canonical kinetic term and slow-roll potential $V(\phi)$ is given by
\begin{eqnarray}\label{action-phi}
S_{\phi} = \int d^3x dt N \sqrt{\gamma} \Big[\, \frac{1}{2N^2} ( \dot{\phi} - N^i\partial_i\phi )^2 
- \frac{1}{2} \gamma^{ij} \partial_i\phi \partial_j\phi - V(\phi) \, \Big] \,.
\end{eqnarray}

We consider the simplest inflationary model so that, similar to the standard matter fields, the inflaton action minimally couples to the gravitational action of the consistent 4DEGB shown in Eq. (\ref{action-EGB}). The total action of the system is given by
\begin{equation}\label{action}
S = S_{\rm g} + S_{\phi} \,.
\end{equation}

In the limit $\tilde{\alpha}\to 0$, the action \eqref{action} reduces to the action of the standard single field inflation and, therefore, we recover the results of the standard slow-roll single field inflation at the background level and any order of perturbations.

\subsection{Background equations}
\label{sec_background}

We consider the spatially flat Friedmann-Lema\^{i}tre-Robertson-Walker (FLRW) background
\begin{equation}\label{metric-BG}
N = \bar{N}(t) \,, \hspace{1cm} N^i = 0 \,, \hspace{1cm} \gamma_{ij} = a(t)^2 \delta_{ij} \,,
\end{equation}
where $a(t)$ is the scale factor. We also consider the homogeneous and isotropic background value $\lambda_{\rm GF} = \bar{\lambda}_{{\rm GF}}(t)$ for the gauge-fixing parameter which after substituting in \eqref{K} results in
\begin{equation}\label{extrinsic-curvature}
{\cal K}^i{}_j = H \delta^i{}_j \,,
\hspace{1cm} H \equiv \frac{{\dot a}}{{\bar N}a} \,,
\end{equation}
where $H$ is the Hubble expansion rate. We see that ${\cal K}^i{}_j$ coincides with the extrinsic curvature of constant time hypersurfaces. This is because the gauge-fixing term does not contribute to the cosmological background equations. Substituting this homogeneous configuration for the gravity sector together with the homogeneous background value $ \phi = \bar{\phi}(t)$ for the scalar field to the total action (\ref{action}), we obtain the minisuperspace action
\begin{equation}\label{action-BG}
{\bar S} = {\cal V} \int dt {\bar N} a^3 \bigg[
\Big( \frac{\dot{\bar{\phi}}}{\bar N}\Big)^2 - V(\phi)
 - 3 M_{\rm Pl}^2 H^2 
- {\tilde \alpha} M_{\rm Pl}^2 H^4 \bigg] \,,
\end{equation}
where ${\cal V}=\int d^3x$ is the spatial volume which will be assumed to be large enough but finite.

Varying the above action with respect to the lapse function, we find the first Friedmann equation
\begin{eqnarray}\label{Friedmann-1}
3 M_{\rm Pl}^2\left( H^2 +\tilde{\alpha} H^4 \right) = \frac{1}{2} \dot{\phi}^2 + V \,,
\end{eqnarray}
where we have set ${\bar N}=1$ after taking the variation by using the time reparametrization symmetry and the dot now coincides with the time derivative with respect to the cosmic time. From now on we set ${\bar N}=1$. Varying the minisuperspace action with respect to the scale factor gives the second Friedmann equation
\begin{equation}\label{Friedmann-2}
-2 M_{\rm Pl}^2 \Gamma \dot{H} = \dot{\phi}^2 \,, \hspace{1cm}  
\Gamma \equiv 1+2{\tilde \alpha}H^2 \,.
\end{equation}

The equation of motion for the scalar field can be deduced by taking variation with respect to the scalar field as 
\begin{equation}\label{KG}
\ddot{\phi} + 3 H \dot{\phi} + V_{,\phi} = 0 \,,
\end{equation}
where $V_{,\phi}$ denotes derivative of the potential with respect to the scalar field.

The geometric slow-roll parameter is given by
\begin{eqnarray}\label{slow-roll}
\epsilon=-\frac{\dot{H}}{H^2} = \epsilon_0 \Big(\frac{\Gamma+1}{2\Gamma}\Big)\,, \hspace{1cm}  
\epsilon_0 \equiv \epsilon|_{{\tilde{\alpha}}=0}  = \frac{3\dot{\phi}^2}{\dot{\phi}^2+2V} \,,
\end{eqnarray}
where $\epsilon_0$ is the slow-roll parameter for the standard inflation in the absence of Gauss-Bonnet corrections. The second geometric slow-roll parameter then turns out to be
\begin{equation}\label{slow-roll-eta}
\eta=\frac{\dot{\epsilon}}{\epsilon{H}} = \eta_0 + \epsilon_0 \Big(\frac{\Gamma-1}{\Gamma^2}\Big)\,, 
\hspace{1cm}  
\eta_0 \equiv \eta|_{{\tilde{\alpha}}=0}= \frac{\dot{\epsilon}_0}{\epsilon_0{H}} \,,
\end{equation}
where $\eta_0$ is the second slow-roll parameter for the standard inflation in the absence of Gauss-Bonnet corrections.

Since $(\Gamma+1)/2\Gamma\leq 1$ and $(\Gamma-1)/\Gamma^2<1$, as far as $\epsilon_0\ll1$ and $\eta_0\ll1$, the slow-roll conditions $\epsilon\ll1$ and $\eta\ll1$ hold irrespective of the values of ${\tilde\alpha}H^2$. Therefore, considering a scalar field with standard slow-roll potential, the quasi-de Sitter solution is always an attractor solution in our setup. In this respect, for a given potential $V(\phi)$ we have one-parameter family of attractor solutions parameterized by the Gauss-Bonnet coupling constant ${\tilde\alpha}$. The Gauss-Bonnet corrections to the background equations are all of the order of ${\tilde \alpha}H^2$ and we do not need to assume ${\tilde \alpha}H^2\ll1$. Indeed as we will show later, depending on the values of ${\tilde{\alpha}}$, ${\tilde \alpha}H^2\gg1$ is as viable as ${\tilde \alpha}H^2\ll1$. We therefore consider the following two interesting regimes
\begin{eqnarray}\label{attractors}
\begin{cases}
\epsilon \approx \epsilon_0 \hspace{.2cm} \mbox{and} \hspace{.2cm} \eta \approx \eta_0 
 &\mbox{for} \hspace{.2cm} {\tilde \alpha}H^2\ll1\\
\epsilon \approx \frac{1}{2} \epsilon_0 \hspace{.2cm} \mbox{and} 
\hspace{.2cm} \eta \approx \eta_0 
\hspace{.5cm} & \mbox{for} \hspace{.2cm} {\tilde \alpha}H^2\gg1
\end{cases}
\end{eqnarray}

For ${\tilde \alpha}H^2\ll1$, we find the standard attractor regime which is expected for ${\tilde \alpha}=0$ while we find a new attractor solution for ${\tilde \alpha}H^2\gg1$ which we call {\it Gauss-Bonnet attractor} since the dominant contribution is given by the Gauss-Bonnet term in this case.

\subsection{Power spectrum of curvature perturbations}

Having shown the existence of an attractor quasi-de Sitter background solution, we consider scalar perturbations around this homogeneous background as follows
\begin{equation}\label{perturbations-scalar}
N = 1 + A \,,\quad  N^i = \delta^{ij} \partial_j B \,, \quad 
\gamma_{ij} = a^2 \big( (1+2 \psi) \delta_{ij} + \partial_i\partial_jE \big) \,, \quad 
\lambda_{\rm GF} = {\bar \lambda}_{\rm GF} + \delta\lambda \quad
\phi = \bar{\phi}(t) + \delta \phi\ \,. 
\end{equation}

We deal with six variables $(A,B,\psi,E,\delta\lambda,\delta\phi)$ for scalar perturbations and the theory is only invariant under the spatial diffeomorphism, which at the level of linear perturbations reduces to $x^i \to x^i + \xi^i$. Using the usual decomposition $\xi^i = \delta^{ij}\partial_j\xi$ in terms of a spatial scalar $\xi$, we can set $E=0$ by fixing this gauge freedom. Moreover, the scalar mode $\delta\lambda$ does not play any roles for the linear scalar perturbations as we have explicitly shown in \cite{Aoki:2020iwm}. We therefore set it to zero $\delta\lambda=0$ and we are left with four scalar modes $(A,B,\psi,\delta\phi)$\footnote{Practically, we can set $\delta \lambda=0$ from the beginning to discuss the dynamics of the curvature perturbations $\zeta$ defined by \eqref{zeta} and to compute the power spectrum and the spectral tilt. However, $\delta \lambda$ is needed to determine the dynamics of all the perturbations variables. See \cite{Aoki:2020iwm} for the detailed discussions.}. Computing the quadratic action from Eq. (\ref{action}), we find that the two variables $A$ and $B$ are non-dynamical and thus can be integrated out. The resultant action apparently has kinetic terms for both of the remaining variables $\psi$ and $\delta\phi$ while we know that there should be only one scalar degree of freedom by definition \cite{Aoki:2020lig}. This fact can be manifestly seen if we define the following combination
\begin{equation}\label{zeta}
\zeta \equiv \psi-\frac{H}{\dot{\bar{\phi}}} \delta \phi \,,
\end{equation}
in terms of which, it is straightforward to show that the quadratic action for the scalar perturbations takes the form \cite{Aoki:2020iwm}
\begin{eqnarray}\label{S2-SS-Fourier}
S^{\rm SS} = M_{\rm Pl}^2 \int dt d^3k\, a^3 \epsilon \Gamma
\left[ \dot{\zeta}^2 -\frac{k^2}{a^2} \zeta^2 \right]
\,,
\end{eqnarray}
where $\Gamma$ is defined in Eq. \eqref{Friedmann-2}. The scalar variable $\zeta$ defined in \eqref{zeta} is nothing but the curvature perturbations which are of interest in inflationary cosmology. The corresponding equation of motion is given by
\begin{align}
\ddot{\zeta} + 3H\left( 1+ \frac{ \eta}{3} 
- \frac{4{\tilde\alpha} \epsilon H^2}{3\Gamma} \right)\dot{\zeta} 
+ \frac{ k^2}{a^2}\zeta = 0  \,, 
\end{align}
which reduces to the standard Mukhanov-Sasaki equation for the single field inflation for ${\tilde\alpha}=0$.

Defining the canonical field
\begin{equation}\label{zeta-canonical}
{\bar \zeta} = M_{\rm Pl} \big( a \sqrt{2\epsilon\Gamma} \big) \zeta \,,
\end{equation}
and working with the conformal time $\tau = \int dt/a(t)$, the action (\ref{S2-SS-Fourier}) becomes
\begin{eqnarray}\label{S2-SS-Fourier-conf}
S^{\rm SS} = \frac{1}{2} \int d\tau d^3k\, 
\left[ {\bar \zeta}'^2 - \bigg(k^2 
- \frac{\big(a \sqrt{\epsilon\Gamma} \big){}''}{a \sqrt{\epsilon\Gamma} } \bigg) {\bar \zeta}^2 \right]
\,,
\end{eqnarray}
where the prime denotes the derivative with respect to the conformal time.

In order to quantize the system, we treat the field ${\bar \zeta}(\tau,{\bf k})$ as a quantum operator and we expand it in terms of creation and annihilation operators as
\begin{eqnarray}\label{nu-op}
\hat{{\bar \zeta}}(\tau,{\bf k})= {\bar \zeta}_{k}(\tau) \hat{a}_{\textbf{k}} 
+ {\bar \zeta}_{k}^{*}(\tau) \hat{a}_{-{\bf k}}^{\dagger} \,, \hspace{1cm}
\big[ a_{\bf k}, a_{{\bf k}'}^{\dagger} \big]= (2\pi)^3 \delta({\bf k}-{\bf k}') \,.
\end{eqnarray}
The mode function ${\bar\zeta}_k(\tau)$ then satisfies
\begin{equation}\label{nu-EoM}
{\bar \zeta}''_k + \bigg( k^2 - \frac{\nu_S^2-1/4}{\tau^2} \bigg) {\bar \zeta}_k = 0 \,,
\end{equation}
where we have defined
\begin{equation}\label{nu-EoM-def}
\nu_S \equiv \frac{3}{2} + \frac{\epsilon}{\Gamma} + \frac{\eta}{2} + \xi_S^2 \,, \hspace{1cm}
\xi_S^2 \equiv \frac{1}{6} \Big(\epsilon\eta+\frac{\dot{\eta}}{H}\Big) 
+ \frac{\epsilon}{3\Gamma} (5\epsilon + \eta) - \frac{2\epsilon^2}{3\Gamma^2} \,.
\end{equation}
We have neglected the terms that are third or higher order in the slow-roll parameters, and also we have used the relation $a=-\frac{1}{(1-\epsilon)H\tau}$ in quasi-de Sitter spacetime. Imposing the Bunch-Davies initial condition, we find the positive frequency solution for Eq. \eqref{nu-EoM} as
\begin{eqnarray}\label{nu-sol-gen}
{\bar \zeta}_k(\tau) = e^{i(1+2\nu_S)\pi/4} \frac{\sqrt{\pi}}{2} \sqrt{-\tau} H^{(1)}_{\nu_S} (-k\tau) \,,
\end{eqnarray}
where $H^{(1)}_{\nu_S}$ is the Hankel function of the first kind and we have chosen the phase factor so that we recover ${\bar \zeta}_k(\tau) = \frac{e^{-ik\tau}}{\sqrt{2k}} \big(1-\frac{i}{k\tau}\big)$ for $\nu_S=3/2$.

The dimensionless power spectrum of the curvature perturbations is defined as $\langle\zeta(\tau,{\bf k})\zeta(\tau,{\bf k}')\rangle\equiv(2\pi^2/k^3)\Delta^2_{\zeta}(k)(2\pi)^3\delta({\bf k}+{\bf k}')$ which for the large scale modes at the time of horizon crossing turns out to be
\begin{equation}\label{PS-zeta}
\Delta^2_{\zeta}(k) = \frac{(1-\epsilon)^2}{8\pi^2\epsilon\Gamma} 
\Big(\frac{H}{M_{\rm Pl}} \Big)^2 \Big(\frac{\Gamma(\nu_S)}{\Gamma(3/2)}\Big)^2
\Big(\frac{-k\tau}{2}\Big)^{3-2\nu_S} \,,
\end{equation}
where $\Gamma(\nu_S)$ is the Gamma function with $\Gamma(3/2) = \sqrt{\pi}/2$. 

The spectral index representing the deviation from the scale-invariance is determined as
\begin{equation}\label{tilt}
n_S -1 \equiv \frac{d\ln\Delta^2_{\zeta}}{d\ln{k}} = 3-2\nu_S 
=  - \frac{2\epsilon}{\Gamma} - \eta - 2 \xi_S^2  \,.
\end{equation}

Now, let us look at the leading term in the power spectrum and the spectral tilt in the two different limits defined in Eq. \eqref{attractors}. Then, from Eq. \eqref{PS-zeta} we find
\begin{equation}\label{PS-zeta-attractors}
\Delta^2_{\zeta} \approx
\begin{cases}
\frac{1}{8\pi^2} \big(\frac{H}{M_{\rm Pl}} \big)^2 \frac{1}{\epsilon_0} 
&\hspace{.5cm}\mbox{for} \hspace{.2cm} {\tilde \alpha}H^2\ll1\\
\frac{1}{8\pi^2}\frac{1}{{\tilde{\alpha}} M_{\rm Pl}^2} \frac{1}{\epsilon_0} 
\hspace{.5cm} 
& \hspace{.5cm} \mbox{for} \hspace{.2cm} {\tilde \alpha}H^2\gg1
\end{cases} \,,
\end{equation}
and from Eq. \eqref{tilt} we find
\begin{equation}\label{tilt-zeta-attractors}
n_S - 1 \approx
\begin{cases} -2\epsilon_0 - \eta_0 
& \hspace{.5cm} \mbox{for} \hspace{.2cm} {\tilde \alpha}H^2\ll1\\ 
- \eta_0
\hspace{.5cm} 
& \hspace{.5cm} \mbox{for} \hspace{.2cm} {\tilde \alpha}H^2\gg1
\end{cases} \,.
\end{equation}

As we see, we find the single field the standard results in the case of standard attractor ${\tilde \alpha}H^2\ll1$.  In the case of the Gauss-Bonnet attractor ${\tilde \alpha}H^2\gg1$, we find new results which we discuss later in Section \ref{sec-obs}.

\section{Gravitational waves}
\label{sec_tensor}

Now we study primordial GWs in this scenario. The tensor perturbations around the background Eq.  (\ref{metric-BG}) are given by
\begin{equation}\label{perturbations-metric-tensor}
N = 1 \,,\quad  N^i = 0 \,, \quad 
\gamma_{ij} = a^2 e^{h_{ij}}  \,, 
\end{equation}
where $h_{ij}$ represents tensor perturbations satisfying the transverse and traceless conditions $\partial^i h_{ij} = 0 = h^{i}{}_{i}$.

Substituting \eqref{perturbations-metric-tensor} in the action \eqref{action} and expanding up to the second order in perturbations, after some integration by parts, we find the quadratic action for the tensor perturbations \cite{Aoki:2020iwm}
\begin{eqnarray}\label{S2-TT-real}
S^{\rm TT} = 
\frac{M_{\rm Pl}^2}{8} \int dt d^3x a^3 \Gamma
\bigg[ \dot{h}_{ij} \dot{h}^{ij} 
- c_T^2 \frac{\partial_k h^{ij} \partial^k h_{ij}}{a^2}
- \frac{\partial^2 h^{ij} \partial^2 h_{ij}}{M^2{a}^4} \bigg] \,,
\end{eqnarray}
which in Fourier space takes the form
\begin{eqnarray}\label{S2-TT-Fourier}
S^{\rm TT} = \frac{M_{\rm Pl}^2}{8} \int dt d^3k a^3 \Gamma 
\bigg[ \dot{h}_{ij} \dot{h}^{ij} - \bigg( c_T^2 \frac{k^2}{a^2} 
+ \frac{1}{M^2} \frac{k^4}{a^4} \bigg) h_{ij} h^{ij}\bigg] \,.
\end{eqnarray}
Here, we have defined
\begin{equation}\label{cs2-T}
c_T^2 \equiv 1 - \epsilon \frac{H^2}{M^2}
\,, \hspace{1cm} 
M \equiv \sqrt{\frac{\Gamma}{4{\tilde\alpha}}}\,,
\end{equation}
$c_T$ denotes the speed of GWs, and $M$ is some mass scale defined by the Hubble expansion rate and the Gauss-Bonnet coupling constant.  

The equation of motion for GWs at the linear order is
\begin{align}\label{GWs-EoM}
\ddot{h}_{ij}+3H \left( 1 - \frac{\epsilon}{3}\frac{H^2}{M^2} \right) \dot{h}_{ij} 
+ \left( c_T^2 \frac{k^2}{a^2} + \frac{1}{M^2} \frac{k^4}{a^4} \right) h_{ij}=0
\,.
\end{align}

\subsection{Dispersion relation and constraints on $\tilde{\alpha}$}

The dispersion relation for the tensor modes can be obtained from the quadratic action \eqref{S2-TT-Fourier}, or equivalently from the equation of motion \eqref{GWs-EoM}, and it takes the form
\begin{equation}\label{DR}
\omega^2 = c_T^2 \frac{k^2}{a^2} + \frac{1}{M^2}\frac{k^4}{a^4} \,.
\end{equation}

From Eqs. \eqref{cs2-T} we see that $c_T^2 \approx 1$ and $c_T^2 \approx 1 - 2\epsilon_0$ for ${\tilde{\alpha}}H^2\ll1$ and ${\tilde{\alpha}}H^2\gg1$ respectively. Therefore, the condition $c_T^2>0$ is always satisfied by the attractor solution in both limits in (\ref{attractors}) as far as the slow-roll conditions hold. Moreover, there also exists the $k^4$ term and its coefficient $M^{-2}$ is positive for both ${\tilde{\alpha}}H^2\ll1$ and ${\tilde{\alpha}}H^2\gg1$ so long as $\tilde{\alpha}>0$. Therefore, the tensor modes are free of either ghost or gradient instabilities for ${\tilde \alpha}>0$ and $\dot{H}<0$.

The appearance of the Lorentz-violating $k^4$ term is the characteristic properties of the theory and it is not properties of the cosmological background geometry that we deal with it here. To see this fact explicitly, we look at the Minkowski limit $\omega^2 = k^2 + 4 {\tilde\alpha} k^4$ where the $k^4$ term is present and dominates at small scales $k\to\infty$ so that $\omega^2 \approx 4 {\tilde\alpha} k^4$. We, therefore, look for the role of this nonlinear term in our model. We first note that for the modes deep inside the horizon, the $k^4$ term dominates and then determines the positive frequency condition for the Bunch-Davies vacuum. Second, since we have computed the power spectrum at the time of horizon crossing, to have observable effects from the $k^4$ term we should estimate its order of magnitude at the time of horizon crossing. In the case of the standard attractor ${\tilde\alpha}H^2\ll1$, the $k^4$ term is suppressed at the time of horizon crossing while we see that it becomes comparable with the standard $k^2$ term for the Gauss-Bonnet attractor ${\tilde\alpha}H^2\gg1$, where we have used $M\approx H/\sqrt{2}$. The Gauss-Bonnet attractor ${\tilde\alpha}H^2\gg1$ then provides a new early universe scenario for the primordial GWs. 

To see whether this regime (${\tilde\alpha}H^2\gg1$) is possible or not, let us look at the current bounds on ${\tilde \alpha}$. We first look at the propagation of the GWs including constraints on the speed of GWs $|1-c_T|\lesssim10^{-15}$ \cite{Monitor:2017mdv}, constraints on the correction that appears in the friction term in Eq. \eqref{GWs-EoM} \cite{Saltas:2014dha,Belgacem:2019pkk}, and constraints on the Lorentz-violating $k^4$ term \cite{Abbott:2017vtc,Sotiriou:2017obf,Gumrukcuoglu:2017ijh}. In Ref. \cite{Aoki:2020iwm}, these bounds are considered and the strongest current bound comes from the Lorentz-violating $k^4$ term which is given by 
\begin{equation}\label{bound}
 {\tilde{\alpha}}\lesssim (10\,\mbox{meV})^{-2} \,.
\end{equation}
Moreover, the theory should reproduce general relativity's prediction in the infrared regime which are well confirmed by the experiments/observations. We note that the theory propagates only two gravitational degrees of freedom and thus there is not any constraint from the fifth force. In this regard, we can only find bounds on ${\tilde \alpha}$ by demanding that corrections coming from the $D\to 4$ Gauss-Bonnet term be smaller than those corresponding to the Einstein-Hilbert term in the infrared regime. In order to do this, following Ref. \cite{Allahyari:2020jkn}, we consider the schematic form for the gravitational action \eqref{action-EGB} as ${\cal R} + {\tilde \alpha}{\cal R}^2$ where ${\mathcal R}$ is an average Gaussian curvature which can be identified by a nonzero tetrad component of the corresponding Riemann tensor. We then should demand ${\tilde \alpha}{\cal R}\lesssim1$ to respect predictions of general relativity in the infrared regime. For a compact astronomical object, the background geometry can be approximated by a Schwarzschild solution and we have ${\cal R} \sim r_{\rm S}/r^3$ on the surface of the compact astronomical object, where $r_{\rm S} = M/(4\pi{M}_{\rm Pl}^2)$ is the corresponding Schwarzschild radius with $M$ and $r$ be mass and radius \cite{Allahyari:2020jkn}. Let us estimate bounds on ${\tilde \alpha}$ for different astronomical objects like sun, neutron star, and black hole shadow by considering a test particle near their surfaces. For sun with mass $M_{\odot}\simeq10^{66}\,{\rm eV}$, and $r_{\odot}\simeq3.5\times10^{15}\,{\rm eV}^{-1}$, we find ${\tilde \alpha}\lesssim10^{36}\,{\rm eV}^{-2}$. For the mercury, the distance from the surface of sun in much larger than the radius of sun and therefore the bound from the precession of mercury would be much weaker. For the shadow of $M87^\ast$ black hole with the mass $M^{87^{*}}_{\rm bh} \simeq 10^{6} M_{\odot}$, demanding that correction to the impact parameter $b$ be small $\delta{b}/b\leq{\cal O}(1)$, we find ${\tilde \alpha}\lesssim10^{32}\,{\rm eV}^{-2}$ \cite{Allahyari:2020jkn}. Finally, for a neutron star with typical values of the mass $M =1.7\times 10^{66}$ eV and radius $r = 5\times10^{10}$ eV$^{-1}$, we find ${\tilde \alpha}\lesssim 10^{22}\,{\rm eV}^{-2}$. We see that the bound from the neutron star is the strongest among astronomical compact objects but still weaker than the bound that we found from the $k^4$ term in the dispersion relation of the GWs in Eq. \eqref{bound}. We also expect stronger bound from the binary black holes which, however, is beyond the scope of the present paper and we leave the investigation of it for a possible future work.

Based on the above discussions, we see that considering the Hubble expansion rate to be less than the Planck scale $H\ll M_{\rm Pl}$, the condition ${\tilde\alpha}H^2\gg1$ can be easily achieved for the ranges of ${\tilde\alpha}$ well below the above bound. Indeed we explicitly confirm this fact in Section \ref{sec-obs} by looking at the CMB observational bounds on the inflationary observables that we computed. Thus, from now on, we focus only on the Gauss-Bonnet attractor ${\tilde{\alpha}}H^2\gg1$ as a new scenario for the primordial GWs.

\subsection{Power spectrum of GWs}

In order to quantize the system we first decompose the tensor modes in terms of the polarization tensors, which satisfy the traceless $e^s_{ii}({\bf k})=0$ and transverse $k^ie^s_{ij}({\bf k})=0$ conditions, as
\begin{equation}\label{tensor-pol}
h_{ij}(\tau,{\bf k}) = \sum_{s} h^s(\tau,{\bf k}) e^s_{ij}({\bf k})\,.
\end{equation}
Defining the canonical field 
\begin{equation}\label{h-canonical}
\bar{h}^s(\tau,{\bf k}) = \frac{M_{\rm Pl}}{2} \big( a \sqrt{\Gamma} \big) h^s(\tau,{\bf k}) \,,
\end{equation}
the action (\ref{S2-TT-Fourier}) in the conformal time takes the form
\begin{eqnarray}\label{S2-TT-Fourier-conf}
S^{\rm TT} = \frac{1}{2} \sum_{s} \int d\tau d^3k\, 
\bigg[ \big| \bar{h'}^s \big|^2 - \bigg( c_T^2 k^2 + \frac{1}{M^2} \frac{k^4}{a^2}
- \frac{\big(a\sqrt{\Gamma}\big){}''}{a\sqrt{\Gamma} } \bigg) \big| \bar{h}^s \big|^2 \bigg]
\,,
\end{eqnarray}
where we have normalized the polarization tensors so that $e^s_{ij}({\bf k}) e^{s'*}_{ij}({\bf k}) = \delta^{ss'}$. Promoting the canonically normalized tensor variables (\ref{h-canonical}) to operators, we expand them in terms of the creation and annihilation operators as
\begin{eqnarray}\label{h-op}
\bar{h}^s(\tau,{\bf k})= \bar{h}_{k}(\tau) a^s_{\textbf{k}} 
+ \bar{h}_{k}^{*}(\tau) a_{-{\bf k}}^{s\dagger} \,; \hspace{1cm}
\big[ a^s_{\bf k}, a_{{\bf k}'}^{s' \dagger} \big]= (2\pi)^3 \delta^{ss'}\delta({\bf k}-{\bf k}') \,.
\end{eqnarray}
The mode function ${\bar h}_k$ satisfies
\begin{equation}\label{h-EoM}
\bar{h}''_k + \bigg( c_T^2k^2 + \frac{H^2}{M^2} k^4 \tau^2 
- \frac{\nu_T^2-1/4}{\tau^2} \bigg) \bar{h}_k = 0 \,,
\end{equation}
where we have defined
\begin{equation}\label{h-op-def}
\nu_T \equiv \frac{3}{2} + \frac{\epsilon}{\Gamma} + \xi_T^2 \,, \hspace{1cm}
\xi_T^2 \equiv -\frac{\epsilon\eta}{3} + \frac{\epsilon}{3\Gamma} (5\epsilon + \eta) 
- \frac{2\epsilon^2}{3\Gamma^2} \,,
\end{equation}
and we have neglected the terms that are third and higher orders in the slow-roll parameters.

The exact solution of Eq. \eqref{h-EoM} can be written in terms of the Whittaker function and its complex conjugate. The positive frequency Bunch-Davies solution is given by (see appendix \ref{app-Whittaker} for the details)
\begin{equation}\label{Whittaker} 
\bar{h}_k = \Big(\frac{M}{H}\Big)^{1/2} \, \frac{e^{-\frac{\pi c_T^2 M}{8H}}}{\sqrt{-2\tau}k} \,
W\left( \frac{ic_T^2 M}{4H}, \frac{\nu_T}{2}, - \frac{iHk^2\tau^2}{M} \right) \,.
\end{equation}
In appendix \ref{app-Whittaker}, we have shown that the above solution correctly recovers the standard result of the linear dispersion relation for $M\gg{H}$. Moreover, for the case of $\nu_T=3/2$, we have shown that our result reduces to the result of Ref. \cite{Ashoorioon:2011eg}, where a nonlinear dispersion relation for the curvature perturbations was studied. Also, we have recovered the results of ghost inflation \cite{ArkaniHamed:2003uz} for $c_T=0$ and  $\nu_T=3/2$.

The two-point correlation function for the tensor helicities then will be
\begin{equation}\label{TPF-h-lambda}
\langle h^s(\tau,{\bf k}) h^{s'}(\tau,{\bf k}') \rangle 
= 4 (1-\epsilon)^2 \frac{H^2 \tau^2}{M_{\rm pl}^2 \Gamma} |\bar{h}_k|^2 \,
\delta^{ss'} \delta({\bf k}+{\bf k}')\,.
\end{equation}
The dimensionless power spectrum is defined as $\langle {h}_{ij}(\tau,{\bf k}) h^{ij}(\tau,{\bf k}')\rangle\equiv(2\pi^2/k^3)\Delta^2_{h}(k)(2\pi)^3\delta({\bf k}+{\bf k}')$. Then relation (\ref{TPF-h-lambda}) leads to 
\begin{equation}\label{PS-h-def}
\Delta^2_{h}(k) = 4 (1-\epsilon)^2 \frac{H^2  k^3 \tau^2}{\pi^2 M_{\rm pl}^2 \Gamma} |\bar{h}_k|^2 \,.
\end{equation}
Substituting Eq. \eqref{Whittaker} in the above relation and then using the small argument behaviour $W(\kappa,\mu,z) \approx \frac{z^{1/2-\mu}\, \Gamma(2\mu)}{\Gamma(\mu + 1/2 - \kappa)}$ for $z\to0$, we find the following result for the limit $k\tau\to0$ 
\begin{equation}\label{PS-h-HC}
\Delta^2_{h}(k) = \frac{2}{\pi^2} (1-\epsilon)^2 
\Big(\frac{H}{M_{\rm Pl}}\Big)^2 \, \frac{1}{\Gamma} \Big(\frac{M}{H}\Big)^{\nu_T}
\, \frac{ \Gamma(\nu_T)^2\, e^{-\frac{\pi c_T^2M}{4H}}}{ \big{|}\Gamma\big(\frac{\nu_T}{2}+\frac{1}{2}-\frac{ic_T^2M}{4H}\big)\big{|}^2}
\big(-k\tau\big)^{3-2\nu_T} \,,
\end{equation}
which also holds at the time of horizon crossing.

The tilt for the tensor power spectrum is given by
\begin{equation}\label{tilt-h}
n_T \equiv \frac{d\ln\Delta^2_{h}}{d\ln{k}} = 3-2\nu_T = - \frac{2\epsilon}{\Gamma} - 2 \xi_T^2 \,.
\end{equation}

Now, neglecting the suppressed slow-roll corrections ($c_T=1$ and $\nu_T=3/2$) and looking at the two different attractor regimes defined in Eq. \eqref{attractors}, from Eq. \eqref{PS-h-HC} we find
\begin{equation}\label{PS-h-attractors}
\Delta^2_{h} \approx
\begin{cases}
\frac{2}{\pi^2} \big(\frac{H}{M_{\rm Pl}} \big)^2
&\mbox{for} \hspace{.2cm} {\tilde \alpha}H^2\ll1 \,\,\, (M\gg H) \\
\frac{c}{\pi^2}\frac{2}{{\tilde{\alpha}} M_{\rm Pl}^2}
\hspace{.5cm} & \mbox{for} \hspace{.2cm} {\tilde \alpha}H^2\gg1
\end{cases}\,,
\end{equation}
where $c \equiv \frac{\pi}{2^{15/4}}\times \frac{e^{-\pi/4\sqrt{2}}}{|\Gamma(1/4-i/4\sqrt{2})|^2}\approx 0.17$ is a numerical constant. In the above, to obtain the standard power spectrum in the limit of ${\tilde \alpha}H^2\ll1$ we have used the relation \eqref{Gamma-identity}. For the spectral tilt of the tensor modes, from Eq. \eqref{tilt-h} we find
\begin{equation}\label{tilt-h-attractors}
n_T \approx
\begin{cases} -2\epsilon_0 
& \hspace{.5cm} \mbox{for} \hspace{.2cm} {\tilde \alpha}H^2\ll1\\ 
\frac{1}{3}\epsilon_0\eta_0
\hspace{.5cm} 
& \hspace{.5cm} \mbox{for} \hspace{.2cm} {\tilde \alpha}H^2\gg1
\end{cases}\,,
\end{equation}
and the tensor to scalar ratio turns out to be
\begin{equation}\label{r-attractors}
r = \begin{cases}
16 \epsilon_0
&\mbox{for} \hspace{.2cm} {\tilde \alpha}H^2\ll1 \,\,\,(M\gg{H})\\
16 c \epsilon_0
\hspace{.5cm} & \mbox{for} \hspace{.2cm} {\tilde \alpha}H^2\gg1
\end{cases}\,.
\end{equation}
We then find modification of the standard relation $r = 16 \epsilon_0$ so that the tensor to scalar ratio is decreased since $c\approx0.17$. Moreover, the consistency condition $r=-8n_T$ is also modified for the Gauss-Bonnet attractor. We discuss these new results in Section \ref{sec-obs}.

\subsection{Order estimate based on scaling dimensions}

Having obtained the power spectrum of the GWs by the direct calculations, it is also useful to look at the dispersion relation by means of which we will be able to estimate the order of magnitude of the power spectrum by the scaling dimensions analysis \cite{ArkaniHamed:2003uz,ArkaniHamed:2003uy,Mukohyama:2009gg}. 

The dispersion relation Eq. (\ref{DR}) includes linear $k^2$ and nonlinear $k^4$ terms. These two terms have different scaling dimensions. We note that when energy scales by a factor $s$ as $E\to{s}E$, the time would scale as $t\to{s}^{-1}t$ by definition. For the nonlinear term we have $\omega \propto k^2$, and therefore we would have $k\to s^{1/2} k$ or equivalently $x\to {s}^{-1/2}x$. Using the fact that quadratic action (\ref{S2-TT-Fourier}) should be invariant under this scaling for the nonlinear term, we find the scaling dimension for the canonical tensor modes as ${\bar h} \to s^{1/4} {\bar h}$ where we schematically represent both polarizations of the tensor modes ${\bar h}^s$ by ${\bar h}$. In this regard, we have ${\bar h} \sim (HM^3)^{1/4}$ for the nonlinear dispersion relation while the similar analysis for the linear dispersion relation leads to the standard result ${\bar h} \sim H$. Now, using the definition Eq. (\ref{h-canonical}), we find amplitude of the standard tensor modes as follows
\begin{equation}\label{h-amplitude0}
h \sim \begin{cases}
\frac{1}{{\tilde\alpha}^{1/2}M_{\rm Pl}} \left(\frac{H}{M}\right)
&\mbox{for} \hspace{.2cm} \omega^2 \propto k^2\\
\frac{1}{{\tilde\alpha}^{1/2}M_{\rm Pl}} \left(\frac{H}{M}\right)^{1/4}
\hspace{.5cm} 
& \mbox{for} \hspace{.2cm} \omega^2 \propto k^4
\end{cases}\,.
\end{equation}
For the standard attractor regime ${\tilde\alpha}H^2\ll1$, from \eqref{cs2-T} we have $M\sim {\tilde\alpha}^{-1/2}$ which after substituting in the above relations, we find the result $h\sim H$ for $k^2$ term and also we see that the effects of $k^4$ term are negligible. Therefore, in the standard attractor regime ${\tilde\alpha}H^2\ll1$ we find the standard power spectrum for the primordial GWs. Of more interest is, however, the Gauss-Bonnet attractor regime ${\tilde\alpha}H^2\gg1$ for which $M \sim H$ and these terms give comparable contributions at the time of horizon crossing as $h\sim {\tilde\alpha}^{-1/2}M_{\rm Pl}^{-1}$. These results are in complete agreement with the result \eqref{PS-h-attractors} which we obtained from the direct calculations. We have therefore confirmed the order of magnitude of the power spectrum without direct computations. From the direct computations we can find the numerical prefactor as we did in \eqref{PS-h-attractors}. The scaling dimension analysis would be also very useful when we estimate the size of non-Gaussianities.

\section{Non-Gaussianity of gravitational waves}\label{sec-NG}

In the previous sections, we have studied the linear perturbations for both the curvature perturbations and GWs. The next step is to study the nonlinear perturbations and find non-Gaussianities (NGs). As it is well known, even in slow-roll single field inflationary models, NG of curvature perturbations is model-dependent \cite{Martin:2013tda}. On the other hand, some properties of NGs of GWs are more or less model independent \cite{Maldacena:2011nz}. Moreover, we have seen that the GWs power spectrum is affected by the novel $k^4$ form of the dispersion relation while we have the standard linear dispersion relation for the curvature perturbations. Therefore, in this section, we look at the NGs of the GWs in a model-independent manner and we leave the analysis of NGs of the curvature perturbations to the future works.

\subsection{Cubic interactions}

Substituting \eqref{perturbations-metric-tensor} in the action \eqref{action} and expanding up to the cubic order, after some integration by parts, we find the cubic action for the tensor modes as follows
\begin{eqnarray}\label{L-hhh}
S^{\rm TTT} =  \sum_{I=1}^4 \int dt \, L^{\rm TTT}_I \,,
\end{eqnarray}
where we have defined the cubic Lagrangians 
\begin{align}\label{L3}
L^{\rm TTT}_1 & \equiv \frac{M_{\rm Pl}^2}{4} a \Gamma c_T^2 
\int d^3x \big( h_{ik} h_{jl} - \frac{1}{2}h_{ij} h_{kl} \big) \partial^i \partial^j h^{kl} \,, \nonumber \\
L^{\rm TTT}_2 & \equiv - \frac{M_{\rm Pl}^2}{a} {\tilde \alpha} 
\int d^3x \bigg( \big( 2 \partial_{j} \partial_{l} h_{ik} - \frac{1}{2}  \partial_{k} \partial_{l}  h_{ij} \big) h^{ij} 
+ \big( h_{ik} h_{jl} - h_{ij} h_{kl} \big) \partial^i \partial^j \bigg) \partial^2h^{kl} \,, \nonumber \\ 
L^{\rm TTT}_3 & \equiv - \frac{M_{\rm Pl}^2}{6} a^3 {\tilde\alpha} H 
\int d^3x \dot{h}^{il} \dot{h}_{ij} \dot{h}^j{}_{l} \,,
\hspace{1.7cm} 
L^{\rm TTT}_4 \equiv \frac{M_{\rm Pl}^2}{2} a {\tilde\alpha} 
\int d^3x \dot{h}^{il} \dot{h}_{ij} \partial^2 h^j{}_{l} \,.
\end{align}

In the limit ${\tilde \alpha}\to0$, the cubic Lagrangian $L^{\rm TTT}_1$ reduces to the standard result in general relativity and all other terms vanish. In the Minkowski limit $H\to0$ (and thus $\Gamma\to 1$ and $c_T\to1$) only cubic Lagrangians $L^{\rm TTT}_1$  and $L^{\rm TTT}_2$ survive and the other two $L^{\rm TTT}_3$  and $L^{\rm TTT}_4$ vanish. Note that $L^{\rm TTT}_2$ shows Lorentz-violating features at the cubic level which survives in the Minkowski limit.

Before performing the direct calculations, let us estimate the order of magnitude of the amplitudes of the three-point functions using the scaling dimension method similar to what we did in the previous section for the two-point function. Using the fact that time derivatives and spatial derivatives are proportional to $\omega$ and $k$, we can find the order of magnitudes of the above interactions as
\begin{align}\label{L3-Orders}
L^{\rm TTT}_1 \sim M_{\rm Pl}^2 h^3 \, \Gamma k^2 \,, \hspace{.5cm}
L^{\rm TTT}_2 \sim M_{\rm Pl}^2 h^3 \, {\tilde\alpha} k^4 \,, \hspace{.5cm}
L^{\rm TTT}_3 \sim M_{\rm Pl}^2 h^3 \, {\tilde\alpha} H \omega^3 \,, \hspace{.5cm}
L^{\rm TTT}_4 \sim M_{\rm Pl}^2 h^3 \, {\tilde\alpha} \omega^2 k^2 \,,
\end{align}
where we have neglected the slow-roll corrections so that $c_T\approx1$. We are interested in the regime ${\tilde\alpha}H^2\gg1$ (where $\Gamma \sim {\tilde\alpha}H^2$) and looking at the dispersion relation \eqref{DR}, we see that all terms have same order of magnitudes at the time of horizon crossing. We therefore need to take into account the effects of all cubic interactions in calculations of the three-point functions.

\subsection{In-in formalism}

Now, we implement the so-called in-in formalism to compute the tensor NG. The dominant contributions from all cubic interactions (\ref{L3}) are given by the simplest case of one-vertex tree-level Feynman diagrams and the three-point correlation function of the tensor polarizations is given by \cite{Maldacena:2002vr,Weinberg:2005vy}
\begin{equation}\label{in-in-formula}
\langle h^{s_1}(\tau,{\bf k}_1) h^{s_2}(\tau,{\bf k}_2) h^{s_3}(\tau,{\bf k}_3) \rangle = 
-i \int_{-\infty}^\tau d\eta \, a(\eta) \big\langle 
\big[h^{s_1}(\tau,{\bf k}_1) h^{s_2}(\tau,{\bf k}_2) h^{s_3}(\tau,{\bf k}_3)
,H_{\rm int}(\eta) \big] \big\rangle \,,
\end{equation}
where $H_{\rm int}$ is the total cubic interaction Hamiltonian given by
\begin{eqnarray}\label{H-int}
H_{\rm int} = \sum_{I=1}^4 H^{(I)}_{\rm int} \,, \hspace{1cm} \mbox{with} \hspace{1cm}
H^{(I)}_{\rm int} = - L^{\rm TTT}_{I} \,,
\end{eqnarray}
and all fields are in the interaction picture. Going to the Fourier space and expanding the tensor modes in terms of the polarization tensors as in \eqref{tensor-pol}, we find the following expressions for the different interaction Hamiltonians
\begin{align}\
H_{\rm int}^{(1)} & = \frac{M_{\rm Pl}^2a \Gamma c_T^2}{4(2\pi)^6} \int d^3p_1 d^3p_2 d^3p_3 
\delta({\bf p}_1+{\bf p}_2+{\bf p}_3) \sum_{s_i} p_3^2\, 
h^{s_1}(\eta,{\bf p}_1) h^{s_2}(\eta,{\bf p}_2) h^{s_3}(\eta,{\bf p}_3) \,
\Pi_1^{s_i}(p_i) \,, \label{H31} \\
H_{\rm int}^{(2)} & =\frac{M_{\rm Pl}^2{\tilde \alpha}}{a(2\pi)^6}
\int d^3p_1 d^3p_2 d^3p_3
\delta({\bf p}_1+{\bf p}_2+{\bf p}_3) \sum_{s_i} p_3^4\,
h^{s_1}(\eta,{\bf p}_1) h^{s_2}(\eta,{\bf p}_2) h^{s_3}(\eta,{\bf p}_3) \,
\Pi_2^{s_i}(p_i) , \label{H32} \\ 
H_{\rm int}^{(3)} & =  \frac{M_{\rm Pl}^2{\tilde\alpha} H}{6(2\pi)^6}  
\int d^3p_1 d^3p_2 d^3p_3
\delta({\bf p}_1+{\bf p}_2+{\bf p}_3) \sum_{s_i}
\frac{d}{d\eta}{h^{s_1}}(\eta,{\bf p}_1) \frac{d}{d\eta}{h^{s_2}}(\eta,{\bf p}_2) 
\frac{d}{d\eta}{h^{s_3}}(\eta,{\bf p}_3) \, \Pi^{s_i}(p_i) , 
\label{H33} \\ 
H_{\rm int}^{(4)} & = \frac{M_{\rm Pl}^2{\tilde\alpha}}{2a(2\pi)^6} \hspace{.1cm}  
\int d^3p_1 d^3p_2 d^3p_3 \delta({\bf p}_1+{\bf p}_2+{\bf p}_3) \sum_{s_i}
p_3^2\, \frac{d}{d\eta}{h^{s_1}}(\eta,{\bf p}_1) \frac{d}{d\eta}{h^{s_2}}(\eta,{\bf p}_2) 
{h^{s_3}}(\eta,{\bf p}_3) \, 
\Pi^{s_i}(p_i) \label{H34}\,,
\end{align}
where we have defined
\begin{align}
&\Pi_1^{s_i}(p_i) \equiv \Pi^{ij}{}_{ik,jl,kl}(s_i,p_i) - \frac{1}{2} \Pi^{ij}{}_{ij,kl,kl}(s_i,p_i) \,, 
\hspace{2cm} \Pi^{s_i}(p_i) \equiv \Pi^{ii}{}_{lk,jl,jk}(s_i,p_i) \,, \nonumber \\
&\Pi_2^{s_i}(p_i) \equiv \Pi^{ij}{}_{jk,il,kl}(s_i,p_i) - \Pi^{ij}{}_{ij,kl,kl}(s_i,p_i) + \frac{p_1^2}{p_3^2} 
\Big[ 2\Pi^{ij}{}_{ik,jl,kl}(s_i,p_i) - \frac{1}{2} \Pi^{ij}{}_{ij,kl,kl}(s_i,p_i) \Big] \,, \label{Pi-I-def}
\end{align}
with
\begin{equation}\label{Pi-def}
\Pi^{ij}{}_{kl,mn,rt}(s_i,{\bf p}_i) \equiv \frac{p_3^i p_3^j}{p_3^2} 
e^{s_1}_{kl}({\bf p}_1) e^{s_2}_{mn}({\bf p}_2) e^{s_3}_{rt}({\bf p}_3) \,.
\end{equation}
Note that although \eqref{Pi-def} depends on the directions of the momenta, the quantities $\Pi_1^{s_i}(p_i)$, $\Pi_2^{s_i}(p_i)$, and $\Pi^{s_i}(p_i)$ which are constructed from the different contractions of the polarizations tensors, depend only on the magnitude of the momenta. The explicit form of these quantities are obtained in appendix \ref{app-polarization} and we use them to find shapes of the bi-spectra later.

\subsection{Amplitude of three-point function}

As usual, we define the amplitude of the three-point function
\begin{equation}\label{A-def}
\langle h^{s_1}(\tau,{\bf k}_1) h^{s_2}(\tau,{\bf k}_2) h^{s_3}(\tau,{\bf k}_3) \rangle \equiv
A^{s_1s_2s_3}(k_1,k_2,k_3) (2\pi)^3 \delta({\bf k}_1+{\bf k}_2+{\bf k}_3) \,,
\end{equation}
and we classify contributions from the different interactions to this amplitude as follows
\begin{equation}\label{Adef}
A^{s_1s_2s_3}(k_1,k_2,k_3) = \sum_{I=1}^4 A^{s_1s_2s_3}_{(I)}(k_1,k_2,k_3) \,.
\end{equation}

Substituting interaction Hamiltonians \eqref{H31}, \eqref{H32}, \eqref{H33}, and \eqref{H34} into the formula \eqref{in-in-formula} and using Wick's theorem, from definitions (\ref{A-def}) and (\ref{Adef}) we find
\begin{align}
A^{s_1s_2s_3}_{(1)} & =  2{\tilde\alpha} M_{\rm Pl}^2 c_T^2 \Big(\frac{M}{H}\Big)^2
\int_{-\infty}^{\tau} \frac{d\eta}{\eta^2}\, {\rm Im}\Big[\prod_{i=1}^3 G_{k_i}(\tau,\eta) \Big] 
k_3^2 \, \Pi_1^{s_1s_2s_3}(k_1,k_2,k_3) + 5 \,\, {\rm perm}
\,, \label{NG-H31-G} \\[10pt]
A^{s_1s_2s_3}_{(2)} & =
2{\tilde\alpha} M_{\rm Pl}^2
\int_{-\infty}^{\tau} d\eta \, {\rm Im}\Big[\prod_{i=1}^3 G_{k_i}(\tau,\eta) \Big] 
k_3^4 \, \Pi_2^{s_1s_2s_3}(k_1,k_2,k_3) + 5 \,\, {\rm perm}\,, 
\label{NG-H32-G} \\[10pt]
A^{s_1s_2s_3}_{(3)} & = 
-\frac{1}{3} \tilde{\alpha} M_{\rm Pl}^2
\int_{-\infty}^{\tau} \frac{d\eta}{\eta} \,
{\rm Im}\Big[\prod_{i=1}^3 \partial_\eta G_{k_i}(\tau,\eta) \Big] 
\Pi^{s_1s_2s_3}(k_1,k_2,k_3) + 5 \,\, {\rm perm}  
\,, \label{NG-H33-G} \\[10pt]
A^{s_1s_2s_3}_{(4)} & =
{\tilde\alpha} M_{\rm Pl}^2
\int_{-\infty}^{\tau} d\eta \, \,
{\rm Im}\Big[ \prod_{i=1}^2\partial_\eta G_{k_i}(\tau,\eta) G_{k_3}(\tau,\eta)
\Big] k_3^2 \, \Pi^{s_1s_2s_3}(k_1,k_2,k_3) + 5 \,\, {\rm perm} 
\,, \label{NG-H34-G}
\end{align}
where we have neglected the slow-roll suppressed corrections and thus set $\nu_T=3/2$, and we have also defined the Wightman function
\begin{eqnarray}\nonumber
G_k(\tau,\eta) \equiv
\frac{4\bar{h}_{k}(\tau) \bar{h}^*_{k}(\eta)}{M_{\rm Pl}^2{\Gamma}a(\tau)a(\eta)}
= \frac{\sqrt{\tau\eta}}{2k^2} \frac{H}{M} 
\frac{e^{-\frac{\pi c_T^2 M}{4H}}}{{\tilde\alpha} M_{\rm Pl}^2} \,
W\left( \frac{ic_T^2 M}{4H}, \frac{3}{4}, -\frac{iHk^2\tau^2}{M} \right)
W\left( -\frac{ic_T^2 M}{4H}, \frac{3}{4}, \frac{iHk^2\eta^2}{M} \right) .
\end{eqnarray}

We are interested in the bi-spectrum after the time of horizon crossing $\tau\to0$ where the Wightman function takes the form
\begin{equation}
G_{k}(0,\eta) = \frac{1}{2{\tilde\alpha}M_{\rm P}^2} \frac{f(k\eta)}{k^3} \,,
\end{equation}
where we have defined
\begin{equation}\label{f}
f(x) \equiv \frac{\sqrt{\pi}}{2} \Big(\frac{H}{M}\Big)^{3/4} \frac{ (-1)^{1/8}e^{-\frac{\pi c_T^2 M}{4H}}}{\Gamma\big(\frac{5}{4}-\frac{ic_T^2M}{4H}\big)} \sqrt{-x} \,
W\left( -\frac{ic_T^2 M}{4H}, \frac{3}{4}, \frac{iH x^2}{M} \right) \,.
\end{equation}

In terms of this new function we find
\begin{align}\
A^{s_1s_2s_3}_{(1)} & = \frac{c_T^2}{(2{\tilde\alpha} M_{\rm Pl}^2)^2} \Big(\frac{M}{H}\Big)^2 \, 
\frac{{\cal I}_{(1)}(x_2,x_3)}{k_1^2k_2^2k_3^2}\, \frac{x_3}{x_2} \, \Pi_1^{s_1s_2s_3}(x_2,x_3) \,
+ 5 \,\, {\rm perm}
\,, \label{NG-H31-s} \\[10pt]
A^{s_1s_2s_3}_{(2)} & =
\frac{1}{(2{\tilde\alpha} M_{\rm Pl}^2)^2} \, 
\frac{ {\cal I}_{(2)}(x_2,x_3)}{k_1^2k_2^2k_3^2}\, \frac{x_3^3}{x_2}\,  \Pi_2^{s_1s_2s_3}(x_2,x_3) \,
+ 5 \,\, {\rm perm} 
\,, \label{NG-H32-s} \\[10pt]
A^{s_1s_2s_3}_{(3)} & = 
\frac{1}{6(2{\tilde\alpha} M_{\rm Pl}^2)^2} \, 
\frac{ {\cal I}_{(3)}(x_2,x_3)}{k_1^2k_2^2k_3^2}\, \frac{1}{x_2x_3}\,  \Pi^{s_1s_2s_3}(x_2,x_3) \,
+ 5 \,\, {\rm perm}
\,, \label{NG-H33-s} \\[10pt]
A^{s_1s_2s_3}_{(4)} & =
\frac{1}{2(2{\tilde\alpha} M_{\rm Pl}^2)^2} \, 
\frac{ {\cal I}_{(4)}(x_2,x_3)}{k_1^2k_2^2k_3^2}\, \frac{x_3}{x_2} \,  \Pi^{s_1s_2s_3}(x_2,x_3) \,
+ 5 \,\, {\rm perm}
\,, \label{NG-H34-s}
\end{align}
where we have defined $x_2\equiv k_2/k_1$, $x_3\equiv k_3/k_1$, and the time integrals
\begin{align}\label{I1}
{\cal I}_{(1)}(x_2,x_3) & = {\rm Im} \Big[ \int_{-\infty}^0 \frac{dy}{y^2} f(y) f(x_2 y) f(x_3 y) \Big] \,, 
\\ \label{I2}
{\cal I}_{(2)}(x_2,x_3) & = {\rm Im} \Big[ \int_{-\infty}^0 dy f(y) f(x_2 y) f(x_3 y) \Big] \,,
\\ \label{I3}
{\cal I}_{(3)}(x_2,x_3) & = {\rm Im} \Big[ -\int_{-\infty}^0 \frac{dy}{y} 
\frac{df(y)}{dy} \frac{df(x_2 y)}{dy} \frac{df(x_3 y)}{dy} \Big] \,,
\\ \label{I4}
{\cal I}_{(4)}(x_2,x_3) & = {\rm Im} \Big[ \int_{-\infty}^0 dy 
\frac{df(y)}{dy} \frac{df(x_2 y)}{dy} f(x_3 y) \Big] \,.
\end{align}

\subsection{Numerical integration method}

\begin{figure}[t] 
\centering
 \includegraphics[width=0.75\linewidth]{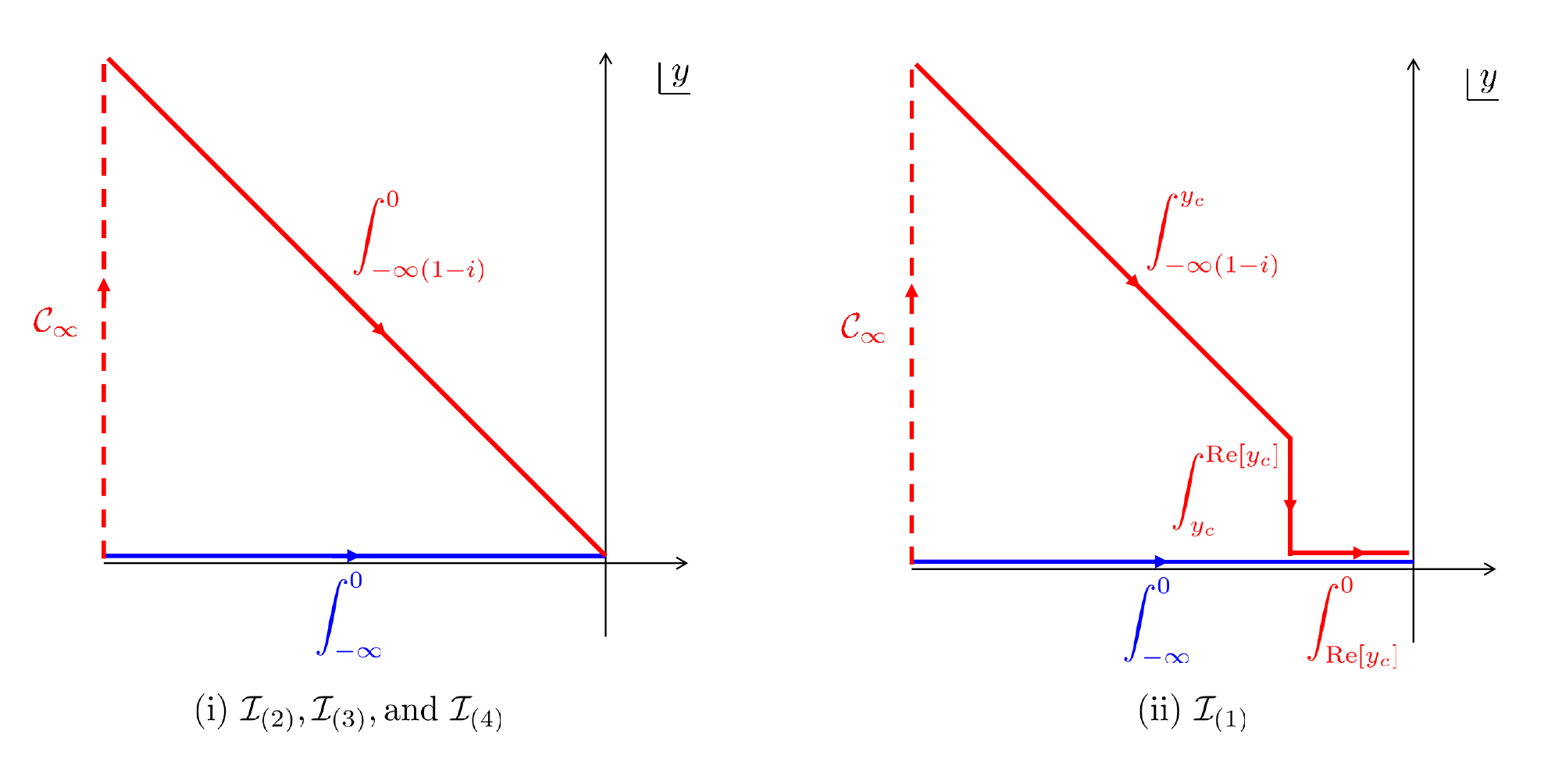}	
 \caption{Deformations of contours. The blue contours are deformed into the red contours where the integral along $\mathcal{C}_{\infty}$ can be ignored because of the asymptotic form \eqref{Whittaker-asymptotic}. }
 \label{fig-contour}
\end{figure}

From the asymptotic form of the Whittaker function in Eq. \eqref{Whittaker-asymptotic}, we see that the integrands in the integrals \eqref{I1}, \eqref{I2}, \eqref{I3}, and \eqref{I4}, oscillate fast in the early times $y\to-\infty$ and the contribution from these oscillations would cancel each other. We, therefore, need to regularize the integrals in this respect and we do it by going to the imaginary plane for $y$ and finding an appropriate contours as shown in Fig.~\ref{fig-contour}. For the integrals \eqref{I2}, \eqref{I3}, and \eqref{I4} we consider the contour $\int_{-\infty(1-i)}^0$ that is shown in subfigure (i) in Fig. \ref{fig-contour} and it works well. However, the real part of the integrand in \eqref{I1} diverges for $y\to0$ in the imaginary plane and we should approach this point on the real line and not imaginary line. In order to do this, we need to change the contour $\int_{-\infty(1-i)}^0$ at a complex time $y_c$ before approaching the time $y=0$. On the other hand, $y_c$ should be chosen late enough so that all contributions from the oscillating terms cancel each other. To find this, we note that we have three oscillating functions in the integrand of integral \eqref{I1} with frequencies given by $1$, $x_2$, and $x_3$. On the other hand, for the bi-spectra amplitudes $A^{s_1s_2s_3}$ we also have permutations of $k_i$ which provide frequencies $x_2^{-1}$ and $x_3^{-1}$. To regularize all oscillating terms, we then choose the complex time $y_c$ as 
\begin{equation}\label{yc}
y_c=\frac{-(1-i)}{{\rm Max}[1,x_2^{-1},x_3^{-1}]} \,,
\end{equation}
and we take contour as $\int_{-\infty(1-i)}^{y_c}+\int_{y_c}^{{\rm Re}[y_c]}+\int_{{\rm Re}[y_c]}^0$ as shown in subfigure (ii) in Fig. \ref{fig-contour}. Note that the second integral is performed parallel to the imaginary line with fixed real value ${\rm Re}[y_c]$ while the last integral is completely performed on the real line. Along these contours we can easily perform the time integrations.

Performing numerically the integrals \eqref{I1}, \eqref{I2}, \eqref{I3}, and \eqref{I4}, following the method that is explained above, we can obtain the bi-spectra for any polarization $s_i$ using the results in \eqref{NG-H31-s}, \eqref{NG-H32-s}, \eqref{NG-H33-s}, and \eqref{NG-H34-s}. However, since our setup is invariant under the parity, we only need to compute $A^{+++}(k_1,k_2,k_3)$ and $A^{++-}(k_1,k_2,k_3)$ and all other amplitudes can be obtained from these quantities by using properties of them under the parity. We also set $c_T\approx1$ by ignoring slow-roll suppressed corrections and set $H/M=\sqrt{2}$ at the time of horizon crossing,

\subsection{Results}

\begin{figure}[ht] 
	\begin{subfigure}[b]{0.5\linewidth}
		\centering
		\includegraphics[width=0.75\linewidth]{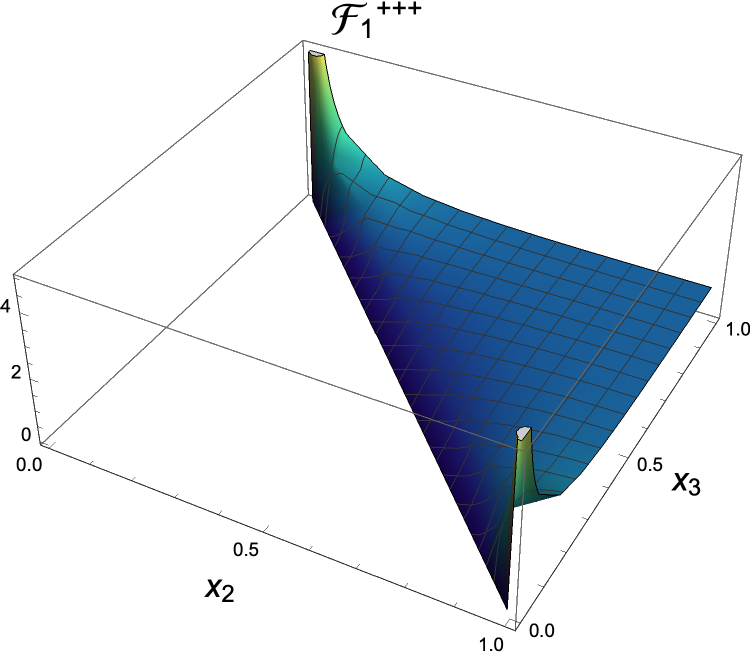}
		\label{fig-F1p} 
		\vspace{4ex}
	\end{subfigure}
	\begin{subfigure}[b]{0.5\linewidth}
		\centering
		\includegraphics[width=0.75\linewidth]{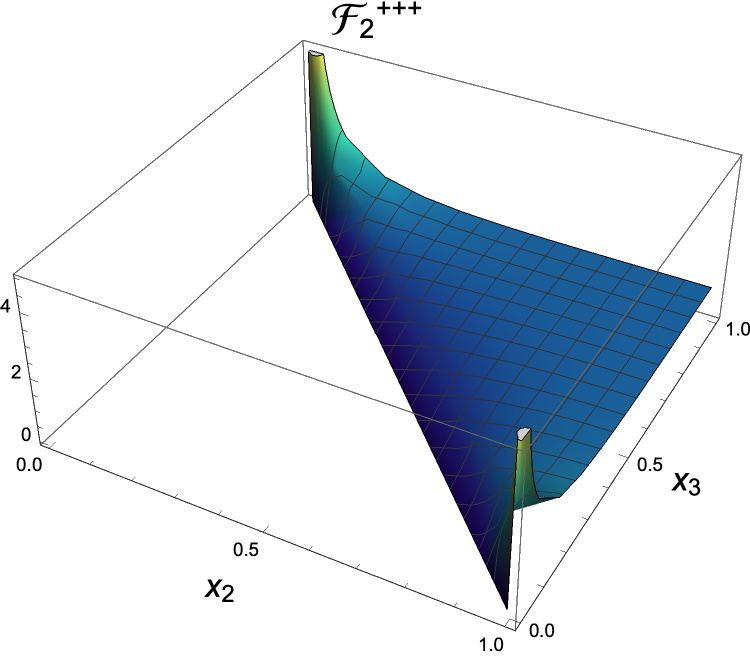} 
		\label{fig-F2p} 
		\vspace{4ex}
	\end{subfigure} 
	\begin{subfigure}[b]{0.5\linewidth}
		\centering
		\includegraphics[width=0.75\linewidth]{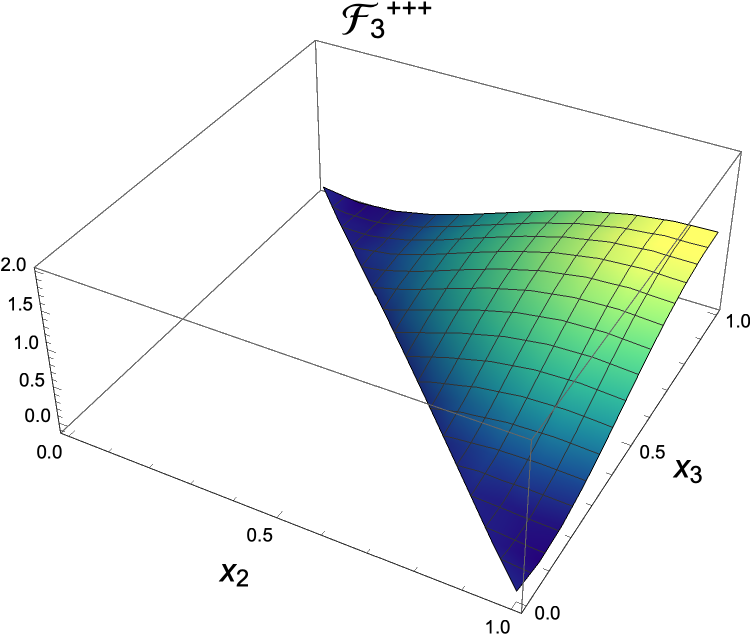} 
		\label{fig-F3p} 
	\end{subfigure}
	\begin{subfigure}[b]{0.5\linewidth}
		\centering
		\includegraphics[width=0.75\linewidth]{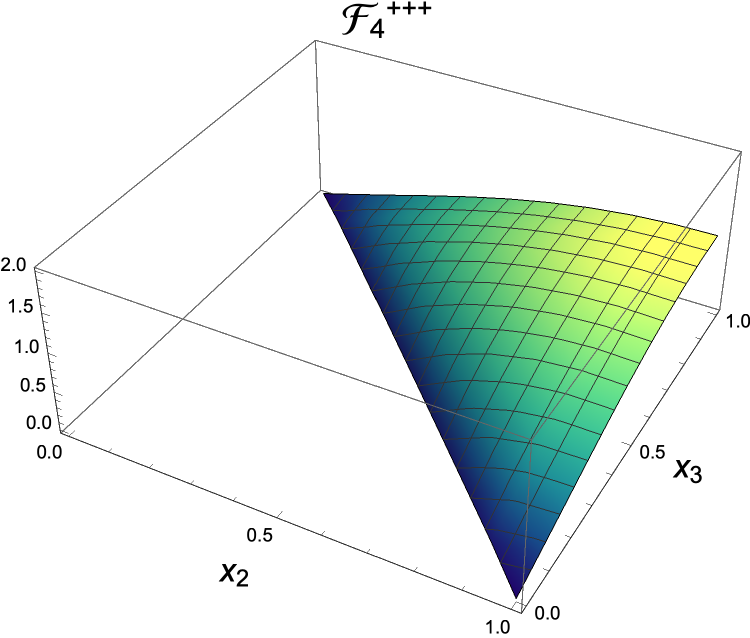} 
		\label{fig-F4p} 
	\end{subfigure} 
\caption{Plots of ${\cal F}^{+++}_{(I)}(x_2,x_3)$ normalized at the equilateral configuration so that ${\cal F}^{+++}_{(I)}(1,1)=1$}
\label{fig-Fp} 
\end{figure}

In the case of $s_1=s_2=s_3=+1$, we find
\begin{equation}\label{NG+++}
A^{+++}_{(I)} = \frac{{\cal A}^{+++}_{(I)}}{(2{\tilde\alpha} M_{\rm Pl}^2)^2} 
\frac{{\cal F}^{+++}_{(I)}(x_2,x_3)}{k_1^2k_2^2k_3^2} \,,
\end{equation}
where ${\cal A}^{+++}_{(I)}$ are given by
\begin{eqnarray}\label{amplitudes+++}
{\cal A}^{+++}_{(1)} \approx 0.516 \,, \hspace{.5cm} {\cal A}^{+++}_{(2)} \approx 1.15 \,, \hspace{.5cm}
{\cal A}^{+++}_{(3)} \approx - 0.02  \,, \hspace{.5cm} {\cal A}^{+++}_{(4)} \approx - 0.073 \,, \hspace{.5cm}
\end{eqnarray}
and ${\cal F}^{+++}_{(I)}(x_2,x_3)$ are normalized with respect to the equilateral configuration as ${\cal F}^{+++}_{(I)}(1,1)=1$. The shape of the NG is completely determined by ${\cal F}^{+++}_{(I)}(x_2,x_3)$ and is shown in Fig. \ref{fig-Fp}.

\begin{figure}[ht] 
	\begin{subfigure}[b]{0.5\linewidth}
		\centering
		\includegraphics[width=0.75\linewidth]{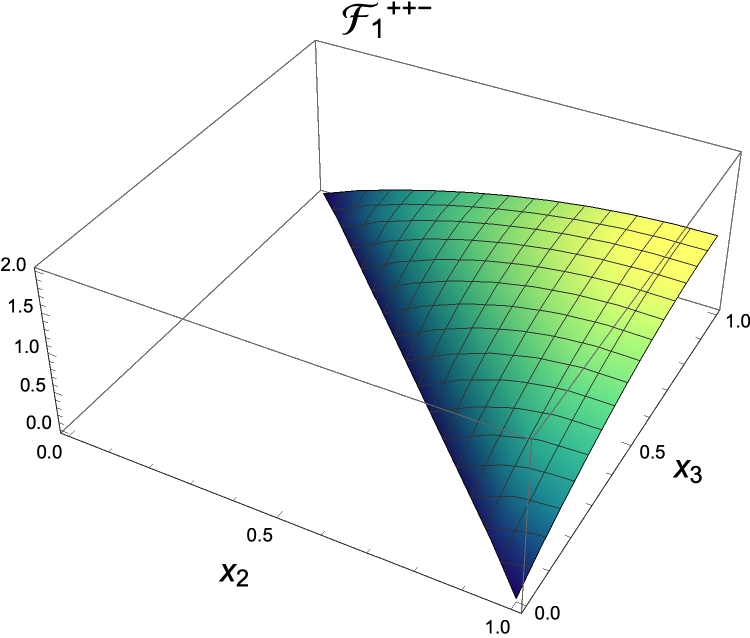}
		\label{fig-F1m} 
		\vspace{4ex}
	\end{subfigure}
	\begin{subfigure}[b]{0.5\linewidth}
		\centering
		\includegraphics[width=0.75\linewidth]{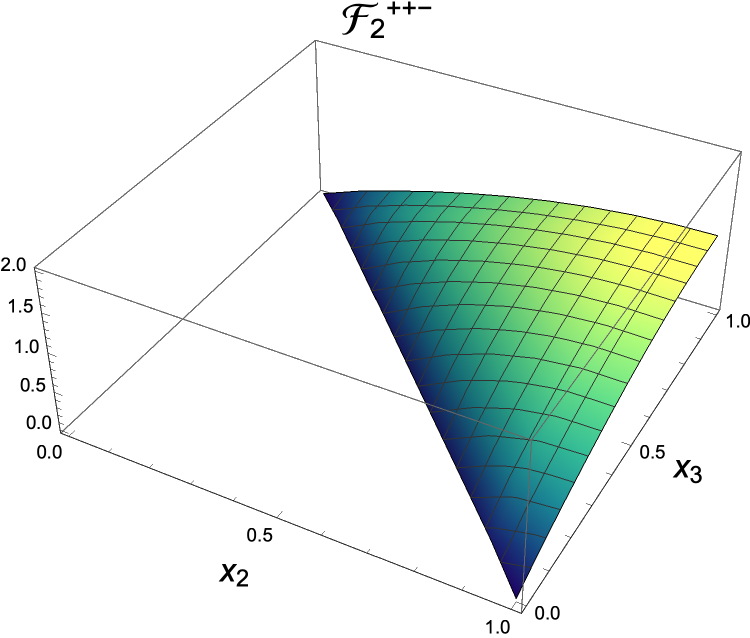} 
		\label{fig-F2m} 
		\vspace{4ex}
	\end{subfigure} 
	\begin{subfigure}[b]{0.5\linewidth}
		\centering
		\includegraphics[width=0.75\linewidth]{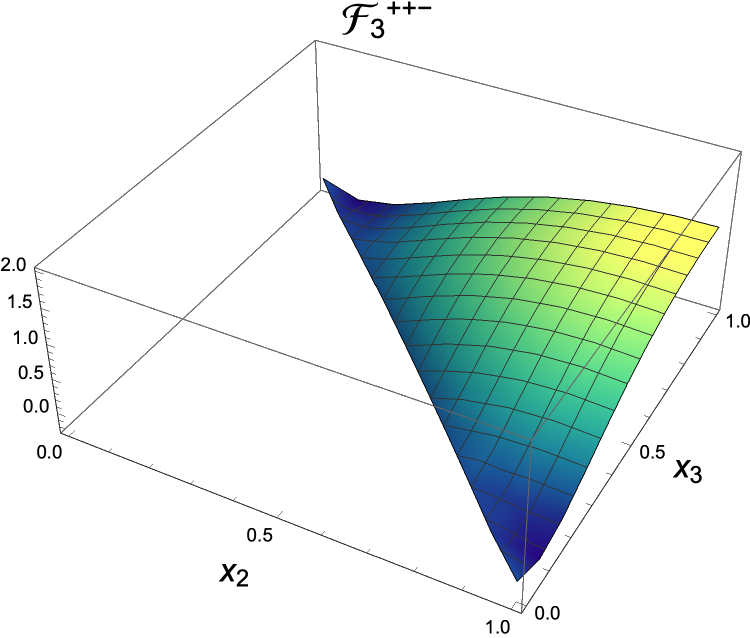} 
		\label{fig-F3m} 
	\end{subfigure}
	\begin{subfigure}[b]{0.5\linewidth}
		\centering
		\includegraphics[width=0.75\linewidth]{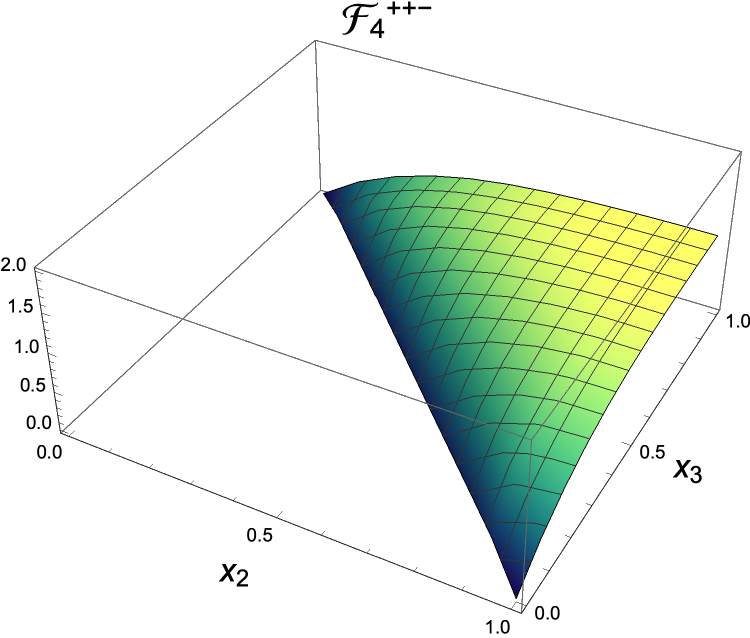} 
		\label{fig-F4m} 
	\end{subfigure} 
	\caption{Plots of ${\cal F}^{++-}_{(I)}(x_2,x_3)$ normalized at the equilateral configuration so that ${\cal F}^{++-}_{(I)}(1,1)=1$}
	\label{fig-Fm} 
\end{figure}

In the case of $s_1=s_2=+1$ and $s_3=-1$, we also find
\begin{equation}\label{NG++-}
A^{++-}_{(I)} = \frac{{\cal A}^{++-}_{(I)}}{(2{\tilde\alpha} M_{\rm Pl}^2)^2} 
\frac{{\cal F}^{++-}_{(I)}(x_2,x_3)}{k_1^2k_2^2k_3^2} \,,
\end{equation}
where ${\cal A}^{++-}_{(I)}$ are given by
\begin{eqnarray}\label{amplitudes++-}
{\cal A}^{++-}_{(1)} \approx 0.058 \,, \hspace{.5cm} {\cal A}^{++-}_{(2)} \approx 0.128 \,, \hspace{.5cm}
{\cal A}^{++-}_{(3)} \approx - 0.002 \,, \hspace{.5cm} {\cal A}^{++-}_{(4)} \approx -0.008 \,, \hspace{.5cm}
\end{eqnarray}
and ${\cal F}^{++-}_{(I)}(x_2,x_3)$ are normalized as ${\cal F}^{++-}_{(I)}(1,1)=1$ and is shown in Fig. \ref{fig-Fm}.

From \eqref{Adef}, we see that the total amplitude of NGs are given by $A^{+++}=\sum_I A^{+++}_{(I)}$ and $A^{++-}=\sum_I A^{++-}_{(I)}$ where $A^{+++}_{(I)}$ and $A^{++-}_{(I)}$ are computed in \eqref{NG+++} and \eqref{NG++-}. Looking at results \eqref{amplitudes+++} and \eqref{amplitudes++-} and also figures \ref{fig-Fp} and \ref{fig-Fm}, we see that the dominant NG is given by the $A^{+++}$ case and has squeezed form. To make it more quantitative, we define the following nonlinear parameters as usual 
\begin{eqnarray}\label{fnl_sq-def}
&&f^{s_1s_2s_3}_{\rm NL,\;sq} \equiv
\lim_{\substack{k_2 \to k_1\\ k_3\to 0}} \frac{A^{s_1s_2s_3}(k_1,k_2,k_3)}{S^{\rm loc}(k_1,k_2,k_3)} \,; \hspace{1cm}
S^{\rm loc}(k_1,k_2,k_3) \equiv \frac{3}{10} (2\pi)^4
\Delta_\zeta^4   \frac{\sum_i k_i^3}{\prod_{i}k_i^3}\,,\\
&&\label{fnl_eq-def}
f^{s_1s_2s_3}_{\rm NL,\;eq} \equiv
\lim_{k_i\to k} \frac{A^{s_1s_2s_3}(k_1,k_2,k_3)}{S^{\rm eq}(k_1,k_2,k_3)} \,;\nonumber\\
&&
S^{\rm eq}(k_1,k_2,k_3) \equiv \frac{9}{10} (2\pi)^4
\Delta_\zeta^4
\left[-\left(\frac{1}{k_1^3 k_2^3} + 2\;{\rm perm} \right)
-\frac{2}{k_1^2 k_2^2 k_3^2} + 
\left(\frac{1}{k_1 k_2^2 k_2^3} + 5\;{\rm perm} \right) \right]\,,
\end{eqnarray}
which quantify the so-called local-type and equilateral-type tensor bispectrum, respectively.

\begin{figure}[t]
\centering
\includegraphics[width=0.5\linewidth]{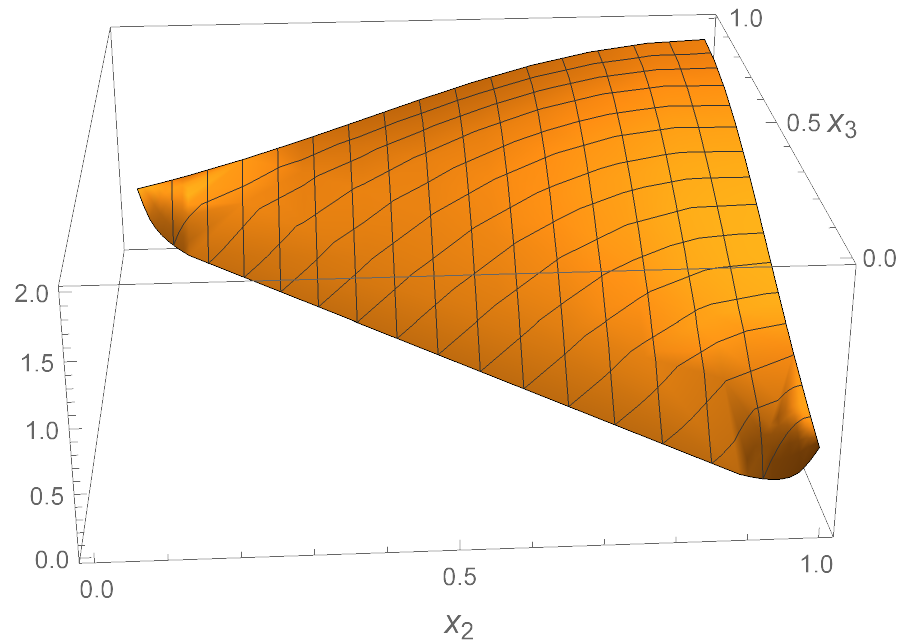}  
\caption{The ratio $\epsilon_0^{-2}A^{+++}/S^{\rm loc}$ is plotted. In the squeezed limit, i.e. $(x_2,x_3)\to (1,0)$ or $\to (0,1)$, the ratio $\epsilon_0^{-2}A^{+++}/S^{\rm loc}$ approaches a constant value.}
\label{fig-fsq}
\end{figure}

In the case of $+++$, from the numerical integration we have seen that it has a peak at the squeezed limit $x_2\to1$ and $x_3\to0$. Adopting the local-type template, we find
\begin{equation}\label{fnl+++}
f^{+++}_{\rm NL,sq} = \frac{5 r^2}{768 c^2} \sum_I \mathcal{A}_{(I)}^{+++}
\lim_{\substack{x_2\to{1}\\x_3\to{0}}} \big[x_3 {\cal F}^{+++}_{(I)}(x_2,x_3)\big]
\approx 0.6 \epsilon_0^2\,,
\end{equation}
where we have computed the limit numerically. In order to do so, we note that both of $A^{+++}$ and $S^{\rm loc}$ are singular at the squeezed limit but must have the same scaling behaviour in this limit. We therefore numerically computed the ratio $A^{+++}/S^{\rm loc}$ as shown in Fig.~\ref{fig-fsq} and we confirmed that the ratio $A^{+++}/S^{\rm loc}$ indeed approaches a constant value in the squeezed limit. The numerical value in \eqref{fnl+++} corresponds to the constant value that is shown in Fig.~\ref{fig-fsq} for the case of $x_2\to1$ and $x_3\to0$.

For the case of $++-$, from the numerical integration we have seen that it has a peak at the equilateral limit 
$x_2\to1$ and $x_3\to1$. Then, adopting the equilateral-type
template, we find
\begin{equation}\label{fnl++-}
f^{++-}_{\rm NL,eq} = \frac{5 r^2}{1152 c^2} \sum_I \mathcal{A}_{(I)}^{++-} \approx 0.195 \epsilon_0^2\,.
\end{equation}

Since the system is invariant under the parity, we have $A^{---}=A^{+++}$ and $A^{--+}=A^{++-}$. Thus, from the above results, we see that for helicities $(+++)$ and $(---)$, $f^{s_1s_2s_3}_{\rm NL,\;sq}$ is larger while $f^{s_1s_2s_3}_{\rm NL,\;eq}$ is larger for helicities $(++-)$ and $(--+)$.

\subsection{Observability}

Let us comment on the observability of the tensor NG generated by this model. Based on the non-linear parameters $f^{ttt}_{\rm NL,sq}$ and $f^{ttt}_{\rm NL,eq}$ that are obtained by summing up all possible tensor helicities $s_1, s_2, s_3$, current constraints within 1$\sigma$ errors
from Planck data \cite{Ade:2015cva,
Akrami:2019izv}
are given by
 \begin{equation}\label{fnl_Planck}
f^{ttt}_{\rm NL,eq} = 600 \pm 1600\,,
\hspace{1cm}
f^{ttt}_{\rm NL,sq} = 290 \pm 180\,.
\end{equation}

From the discussion on the future prospects, although the error is still $\Delta f^{ttt}_{\rm NL} \sim \mathcal{O} (1)$ for the on-going project like LiteBIRD \cite{Hazumi:2012gjy},  we can expect that $\Delta f^{ttt}_{\rm NL, sq}$ may be improved up to $ \sim \mathcal{O} (10^{-4})$ \cite{Shiraishi:2019yux}, which is at the same level of the theoretical prediction of this model (see Eq.~(\ref{fnl+++})).

\section{Constraints on the parameters from CMB data}\label{sec-obs}

In the previous sections, we have seen that the scalar and tensor power spectra reduce to their standard counterpart in the limit ${\tilde\alpha}H^2\ll1$ while they are different for the Gauss-Bonnet attractor limit ${\tilde\alpha}H^2\gg1$. In this section we look at the observational bounds on the amplitude and tilts of the power spectra in this limit and discuss implications of the model accordingly.

The amplitude and the tilt of the scalar power spectrum for the Gauss-Bonnet attractor ${\tilde\alpha}H^2\gg1$ are given in \eqref{PS-zeta-attractors} and \eqref{tilt-zeta-attractors} as
\begin{equation}\label{PS-zeta-GB}
\Delta^2_{\zeta} \approx
\frac{1}{8\pi^2}\frac{1}{{\tilde{\alpha}} M_{\rm Pl}^2} \frac{1}{\epsilon_0} \,, \hspace{1cm}
n_S - 1 \approx - \eta_0 \,.
\end{equation}
For the GWs from \eqref{PS-h-attractors} and \eqref{tilt-h-attractors} we have
\begin{equation}\label{PS-h-GB}
\Delta^2_{h} \approx
\frac{c}{\pi^2}\frac{2}{{\tilde{\alpha}} M_{\rm Pl}^2} \,, \hspace{1cm}
n_T \approx \frac{1}{3}\epsilon_0\eta_0 \,,
\end{equation}
where $c = 0.17$ is a numerical constant. The tensor to scalar ratio is then 
\begin{equation}\label{r-GB}
r = 16 c \epsilon_0 \,.
\end{equation}

Some comments are in order. First, contrary to the standard slow-roll inflation, the spectral tilt of curvature perturbations does not depend on $\epsilon_0$ and, as a result, the observational constraint on the tilt requires a small positive $\eta_0$. Also, the tensor-to-scalar ratio is reduced from the standard case by a constant numerical factor $c\approx 0.17$. Therefore the CMB constraint on the inflaton action needs to be revisited. Second, we see that the tilt of tensor power spectrum is second order in slow-roll parameters and therefore we have almost scale-invariant primordial GWs. Third, both the scalar power spectrum \eqref{PS-zeta-GB} and tensor power spectrum \eqref{PS-h-GB} do not depend on the Hubble parameter which means that the scale of inflation cannot be fixed by the CMB observations. Indeed, we only need to satisfy the condition ${\tilde\alpha}H^2\gg1$ to get the Gauss-Bonnet attractor. From the CMB observations, then we only get a lower bound $H\gg{\tilde\alpha}^{-1/2}$. Taking $\Delta^2_\zeta\sim10^{-9}$, $n_S\sim0.96$, $r\sim10^{-1}$, we find $\epsilon_0\sim\eta_0\sim 10^{-2}$ and
\begin{equation}\label{alpha-CMB}
{\tilde\alpha} \sim (10^{14} \, \mbox{GeV})^{-2}=10^{-46} \, \mbox{eV}^{-2} \,,
\end{equation}
which is much below the current upper bound we already pointed out in Eq. \eqref{bound}. Therefore the assumption ${\tilde\alpha}H^2\gg1$ that we considered from the beginning can be safely satisfied. From the above value we find the following lower bound for the Hubble parameter
\begin{equation}\label{bound-H}
H \gg 10^{14} \, \mbox{GeV} \,,
\end{equation}
to guaranty the Gauss-Bonnet attractor condition ${\tilde\alpha}H^2\gg1$. We see that the scale of inflation in our model would be higher than the standard one $H \sim 10^{14}\, \mbox{GeV}$. Thus, contrary to the standard slow-roll single field inflation, detecting primordial GWs cannot fix the scale of the inflation in our scenario for the Gauss-Bonnet attractor ${\tilde\alpha}H^2\gg1$.

Physically, the $H$-independence of the power spectra may be understood by the scale invariance of the Gauss-Bonnet attractor. Since the Gauss-Bonnet term $R^2_{\rm 4DGB}$ is 4th-order in derivatives, the 4D Gauss-Bonnet term is globally scale invariant. In the standard attractor $\tilde{\alpha}H^2 \ll 1$, the power spectra of the scalar and tensor perturbations increase as the height of the inflaton potential increases. Once the Gauss-Bonnet contributions dominate, the overall scale of the potential must be irrelevant because of the scale invariance, meaning that $\tilde{\alpha}^{-1}$ should determine the maximum scale of the power spectra. Note that the scale invariance is the global one. Therefore, the derivatives of the potential have the physical meaning even in the Gauss-Bonnet attractor and the $\epsilon_0$-dependence of the scalar power spectrum should remain as we have confirmed explicitly. Then, the scalar power spectrum has the tilt at the first order in the slow-roll parameters. On the other hand, the tensor power spectrum is determined only by $H^2$ in the standard attractor and $H^2$ should be replaced with $\tilde{\alpha}^{-1}$ in the Gauss-Bonnet attractor. As result, the tensor power spectrum has no tilt in the Gauss-Bonnet attractor at the first order in the slow-roll parameters.

Despite these distinctions between the scalar power spectrum and the tensor power spectrum, it is worth mentioning that from Eqs. \eqref{PS-zeta-GB}, \eqref{PS-h-GB}, and \eqref{r-GB}, we find a consistency relation for our model
\begin{equation}\label{CR}
n_T = -\frac{r}{8} \, \frac{n_S-1}{6c} \,,
\end{equation}
which is different than the standard one  where tilt of tensor power spectrum is only related to the tensor to scalar ratio through the relation $n_T=-r/8$. The inflationary models can be also classified based on the relations between $n_T$, $n_S$, and $r$ \cite{Bassett:2005xm} and, therefore, by measuring $n_T$ in future, the new consistency relation \eqref{CR} can be testified by the observations\footnote{Deviation from the standard consistency relation also happens in the single field models with non-trivial sound speed \cite{Garriga:1999vw}. As for $k$-inflation, the consistency relation is given by $n_T=-\frac{r}{8} \frac{1}{c_S}$, where $c_S$ is the sound speed of the scalar perturbations.}.

\section{Summary and discussions}\label{summary}

In this paper we have studied slow-roll single field inflationary universe in the context of the recently proposed consistent $D\to4$ Einstein-Gauss-Bonnet gravity. Assuming the standard slow-roll conditions for the inflaton, we have found one-parameter family of the attractor solutions labeled by the rescaled Gauss-Bonnet coupling constant ${\tilde\alpha}$. For ${\tilde\alpha}H^2\ll1$ (or equivalently ${\tilde\alpha}\to0$), we find the standard slow-roll inflation. On the other hand, for ${\tilde\alpha}H^2\gg1$, we find a new attractor regime which we called the Gauss-Bonnet attractor. We have studied linear scalar perturbations around this new Gauss-Bonnet attractor and computed the power spectrum of the curvature perturbations and the associated spectral tilt. In the tensor sector at the linear level, the dispersion relation for the GWs takes nonlinear $k^4$ correction from the Gauss-Bonnet term. We have shown that the effects of the nonlinear $k^4$ term on observable quantities have the same order of magnitude as those of the standard linear $k^2$ term for the Gauss-Bonnet attractor ${\tilde\alpha}H^2\gg1$, while they are suppressed for the standard attractor regime ${\tilde\alpha}H^2\ll1$. We have therefore found a new scenario for the GWs. Around the Gauss-Bonnet attractor we have found that the power spectra of curvature perturbations and GWs do not depend on the Hubble expansion rate and therefore the scale of inflation cannot be fixed by the CMB observations. We have only found a lower bound for the Hubble parameter for Gauss-Bonnet attractor so that inflation with $\tilde{\alpha}H^2\gg 1$ would happen at scales comparable to or higher than that of the standard slow-roll single field inflation $H\sim10^{14}$ GeV. We have then found a model-independent relation and bounds from the CMB observational data. From the observed red-tilt spectrum for the curvature perturbations, we have found $\eta_0\sim10^{-2}$. The tilt of GWs is second order in the slow-roll parameters and we thus have almost scale-invariant GWs in this scenario. Finally, we studied NGs of GWs in this model and we estimated the nonlinear parameters $f^{s_1s_2s_3}_{\rm NL,\;sq}$ and $f^{s_1s_2s_3}_{\rm NL,\;eq}$ by fitting the computed GWs bispectra with the local-type and equilateral-type templates respectively at the squeezed limit and at the equilateral shape. We have shown that for helicities $(+++)$ and $(---)$, $f^{s_1s_2s_3}_{\rm NL,\;sq}$ is larger while $f^{s_1s_2s_3}_{\rm NL,\;eq}$ is larger for helicities $(++-)$ and $(--+)$. The new inflationary scenario that we have studied here can be testified by the CMB observations and with future more precise data can be discriminated from the standard slow-roll single field inflation.

\vspace{0.7cm}

{\bf Acknowledgments:} 
M.A.G. thanks Seyed Ali Hosseini Mansoori for insightful discussions. K.A. and M.A.G. acknowledge the xTras package~\cite{Nutma:2013zea} which was used for tensorial calculations. The work of K.A. was supported in part by Grants-in-Aid from the Scientific Research Fund of the Japan Society for the Promotion of Science, No.~19J00895 and No.~20K14468. The work of M.A.G. was supported by Japan Society for the Promotion of Science Grants-in-Aid for international research fellow No. 19F19313. 
The work of S.Mi was supported in part by Japan Society for the Promotion of Science Grants-in-Aid for Scientific Research No.~20H04749.
The work of S.Mu was supported in part by Japan Society for the Promotion of Science Grants-in-Aid for Scientific Research No.~17H02890, No.~17H06359, and by World Premier International Research Center Initiative, MEXT, Japan. 

\vspace{0.7cm}

\appendix

\section{Whittaker mode function and its asymptotic behaviours}\label{app-Whittaker}
\setcounter{equation}{0}
\renewcommand{\theequation}{A\arabic{equation}}

In this appendix, we present the details of the Bunch-Davies solution of Eq. \eqref{h-EoM}
\begin{equation}\label{h-EoM-app}
\bar{h}''_k + \bigg( c_T^2k^2 + \varepsilon^2 k^4 \tau^2 
- \frac{\nu_T^2-1/4}{\tau^2} \bigg) \bar{h}_k = 0 \,; \hspace{1cm} \varepsilon \equiv \frac{H}{M} \,,
\end{equation}
and we then discuss the different limits of the solution to recover the well-known solutions in the literature. The case with $\varepsilon\to0$ (or equivalently $M\gg{H}$) corresponds to the standard case with linear dispersion relations while $c_T\to0$ corresponds to the case of ghost inflation \cite{ArkaniHamed:2003uz}. In this paper, we are interested in the regime where neither $\varepsilon$ nor $c_T$ are small.

Working with the following variables
\begin{equation}\label{Eq-CHV}
{\bar h}_k(z) = c_1 \frac{w_{k}(z)}{(\varepsilon{z})^{1/4}} \,, 
\hspace{1cm} 
z \equiv -i\varepsilon k^2 \tau^2 \,, 
\end{equation}
from Eq. \eqref{h-EoM-app}, we find that $w_{k}(z)$ satisfies the equation of motion
\begin{equation}\label{h-EoM-app-CHV}
w''_k + \bigg( -\frac{1}{4} + \frac{\kappa}{z} + \frac{1/4-\mu^2}{z^2} \bigg) w_k = 0 \,; 
\hspace{1cm} 
\mu \equiv \frac{\nu_T}{2} \,, \hspace{.5cm} \kappa \equiv \frac{ic_T^2}{4\varepsilon} \,.
\end{equation}

The general solution for $w_k(z)$ is given by the linear combination of the Whittaker functions $W(\kappa,\mu,z)$ and $M(\kappa,\mu,z)$ \cite{Abramowitz:1972}. For large argument limit $z\to\infty$, Whittaker functions have the following asymptotic behaviors
\begin{equation}\label{Whittaker-asymptotic}
W(\kappa,\mu,z) \sim e^{-z/2} z^{\kappa} \,, \hspace{1cm}
M(\kappa,\mu,z) \sim \frac{\Gamma(1+2\mu)}{1/2+\mu-\kappa} e^{z/2} z^{-\kappa} \,.
\end{equation}

Looking at the definition of $z$ in Eq. \eqref{Eq-CHV}, we see that for the modes deep inside the horizon $-k\tau\to\infty$, the Whittaker function $W(\kappa,\mu,z)$ provides positive frequency solution and, therefore, we only keep this branch of solution for the Bunch-Davies vacuum. In this regard, the positive frequency solution for Eq. \eqref{h-EoM-app} is given by
\begin{equation}\label{sol-WF-c1}
{\bar h}_k(z) = c_1 \frac{W(\kappa,\mu,z)}{(\varepsilon{z})^{1/4}} \,.
\end{equation}

The remaining task is to find the constant $c_1$ through the Wronskian condition 
\begin{equation}
{\bar h}(\tau) \frac{d{\bar h}^*(\tau)}{d\tau} - \frac{d{\bar h}(\tau)}{d\tau} {\bar h}^*(\tau) = i \,,
\end{equation}
which is guaranteed by the commutation relations in Eq. \eqref{h-op}. Rewriting the above conditions in terms of $w_k(z)$ with $z$ defined in \eqref{Eq-CHV} and using the fact that $W^*(\kappa,\mu,z) = W(\kappa^*,\mu^*,z^*) = W(-\kappa,\mu,-z)$, where the last step only holds in our special case, and also using the useful relation $W(\kappa,\mu,z)dW(-\kappa,\mu,-z)/dz-dW(\kappa,\mu,z)/dz\, W(-\kappa,\mu,-z) = e^{-i\pi\kappa}$, we find the following solution
\begin{equation}\label{sol-WF}
{\bar h}_k(z) = e^{\frac{i\pi}{8}} \frac{e^{-\frac{\pi{c}_T^2}{8\varepsilon}}}{\sqrt{2k}} \frac{W(\kappa,\mu,z)}{(\varepsilon{z})^{1/4}} \,.
\end{equation}

Neglecting the constant phase factor which does not contribute to the cosmological correlation functions in which we are interested as observable quantities and substituting Eq. \eqref{Eq-CHV} and Eq.  \eqref{h-EoM-app-CHV} for $z$, $\kappa$, and $\mu$ in the above solution we find the Bunch-Davies solution Eq. \eqref{Whittaker} which we use to compute power spectrum and NGs in this paper.

The wave function \eqref{sol-WF} correctly reduces to the result of Ref. \cite{Ashoorioon:2011eg} for $\nu_T=3/2$ where the effects of nonlinear dispersion relation for the curvature perturbations is studied. Let us look at the limit $c_T\to0$ which corresponds to the case of the dispersion relation for the scalar mode in ghost inflation \cite{ArkaniHamed:2003uz}. In this case using the following identity
\begin{equation}
W(0,\mu,z) = \sqrt{\frac{z}{\pi}} K_\mu\left(\frac{z}{2}\right) = \frac{\sqrt{\pi{z}}}{2} i^{\mu+1} H^{(1)}_{\mu}\left(\frac{iz}{2}\right)\,,
\end{equation}
where $K_\mu$ is the modified Bessel function of the second kind and $H^{(1)}_{\mu}=J_\mu+iY_\mu$ is the Hankel function of the first kind with $J_\mu$ and $Y_\mu$ being the standard Bessel functions of the first and second kinds respectively, we find 
\begin{equation}\label{MF-ghost}
\bar{h}_k = e^{-\frac{3i\pi}{8}} \sqrt{\frac{\pi}{8}} \sqrt{-\tau} H^{(1)}_{3/4} \left(\frac{Hk^2\tau^2}{2M}\right) \,,
\end{equation}
which up to a constant phase factor coincides with the result of Ref. \cite{ArkaniHamed:2003uz}.

Now, we look at the limit $\varepsilon\to0$ (or equivalently $M\gg{H}$) which corresponds to the standard case with linear dispersion relation. In order to do so, we write the Whittaker function in terms of the Kummer function $U$ through the identity
\begin{equation}\label{W-U}
W(\kappa,\mu,z) = e^{-z/2} z^{\mu+1/2} U(1/2+\mu-\kappa,1+2\mu,z) \,.
\end{equation}
From Eq. \eqref{h-EoM-app-CHV}, we see that limit $\varepsilon\to0$ corresponds to $\kappa\to\infty$ and in this limit we can use the following relation \cite{Abramowitz:1972}
\begin{equation}
\lim_{a\to\infty} \Big[ \Gamma(1+a-b) U(a,b,-z/a) \Big] = - i \pi e^{i\pi {b}} z^{1/2-b/2} H^{(1)}_{b-1}(2\sqrt{z}) \,,
\end{equation}
which gives
\begin{equation}\label{W-limit-1}
\lim_{\varepsilon\to0} W(\kappa,\mu,z) = 
\pi e^{i\frac{\pi}{2}(1+4\mu)} 
\lim_{\varepsilon\to0} \Big[\frac{z^{1/2} \kappa^{-\mu}}{\Gamma(1/2-\mu-\kappa)} H^{(1)}_{2\mu}(2\sqrt{\kappa{z}})\Big] \,.
\end{equation}
From Eq. \eqref{h-EoM-app-CHV} we have $z=|{\rm Im}[z]|e^{i\pi/2}$ and $\kappa=|{\rm Im}[\kappa]|$. We then use the following asymptotic identity
\begin{equation}\label{Gamma-identity}
\lim_{|a|\to\infty} |\Gamma(b+ia)|= \sqrt{2\pi} |a|^{b-1/2}e^{-\pi|a|/2} \,,
\end{equation}
which is valid for finite values of $b$ and find the result
\begin{equation}\label{W-limit-final}
\lim_{\varepsilon\to0} W(\kappa,\mu,z) = 
e^{i\theta} \sqrt{\frac{\pi}{2}} |{\rm Im}[z]|^{1/2} e^{\pi |{\rm Im}[\kappa]|/2}
H^{(1)}_{2\mu}\big(2\sqrt{|{\rm Im}[z]{\rm Im}[\kappa]|}\big) \,,
\end{equation}
where we have defined the phase $\theta\equiv 3\pi/4+2\mu+{\rm Im}[\Gamma(1/2-\mu-\kappa)]/{\rm Re}[\Gamma(1/2-\mu-\kappa)]$.
Using Eqs. \eqref{W-limit-final} in \eqref{sol-WF}, we finally find the well-known result
\begin{equation}\label{sol-WF-standard}
\lim_{\varepsilon\to0} {\bar h}_k(z) = e^{i\theta} \frac{\sqrt{\pi}}{2} 
\sqrt{-\tau} H^{(1)}_{\nu_T}(-c_Tk\tau) \,.
\end{equation}

In this appendix and throughout the whole paper we worked with the principal values $-\frac{\pi}{2}\le {\rm arg}[z]\le +\frac{\pi}{2}$ for a complex variable $z$.

\section{Circular polarization tensors}\label{app-polarization}
\setcounter{equation}{0}
\renewcommand{\theequation}{B\arabic{equation}}

In this appendix, we compute explicit expressions for the terms with different contractions of momenta and polarizations tensors which appear in calculations of the GWs three-point correlation functions.

From the conservation of momentum, we always deal with the case of ${\bf p}_1+{\bf p}_2+{\bf p}_3=0$. Therefore, the momenta ${\bf p}_i$ are restricted to a plane and the matrix ${\bf e}^{s_i}({\bf p}_i)$ made of components of the circular polarization tensor $e^{s_i}_{ij}({\bf p}_i)$ simplifies to \cite{McFadden:2011kk,Soda:2011am}
\begin{eqnarray}\label{eij}
{\bf e}^{s_i}({\bf p}_i) := \frac{1}{2}
\begin{pmatrix}
\sin^2\varphi_i & - \sin\varphi_i \cos\varphi_i & -i s_i \sin\varphi_i\\
-\sin\varphi_i \cos\varphi_i & \cos^2\varphi_i & i s_i \cos\varphi_i\\
-i s_i \sin\varphi_i & i s_i \cos\varphi_i & -1
\end{pmatrix} \,,
\end{eqnarray}
where $\varphi_i$ are azimuthal angles which determine ${\bf p}_i$. Every momenta has its own magnitudes while their direction can be completely fixed through two relative angles $\varphi_2-\varphi_1$ and $\varphi_3-\varphi_1$. Therefore, without loss of generality, we choose the origin so that $\varphi_1=0$ and the conservation of momentum yields 
\begin{eqnarray}\label{phi-i}
&&\cos \varphi_2 = \frac{x_3^2-x_2^2-1}{2x_2}\,,
\hspace{1cm}
\sin \varphi_2 = \frac{\lambda}{2 x_2}\,,
\hspace{1cm}\cos \varphi_3 = \frac{x_2^2-x_3^2-1}{2x_3}\,,
\hspace{1cm}
\sin \varphi_3 = -\frac{\lambda}{2 x_3}\,,\nonumber\\
&&\hspace{1cm} \mbox{with} 
\hspace{1cm} 
\lambda=\sqrt{2 x_2^2 + 2 x_2^2 x_3^2 + 2 x_3^2 -1-x_2^4 -x_3^4}\,,
\end{eqnarray}
where we have defined the wave number ratios
\begin{equation}\label{p-ratio}
x_{2}\equiv \frac{p_2}{p_1 }\,,\hspace{1cm} x_{3}\equiv \frac{p_3}{p_1} \,.
\end{equation}

Now, we define the following tensor 
\begin{equation}\label{Pi-def-app}
\Pi^{ij}{}_{kl,mn,rt}(s_i,{\bf p}_i) \equiv \frac{p_3^i p_3^j}{p_3^2} 
e^{s_1}_{kl}({\bf p}_1) e^{s_2}_{mn}({\bf p}_2) e^{s_3}_{rt}({\bf p}_3) \,,
\end{equation}
from which we can define all the quantities that we need in the calculations of the three-point functions as follows
\begin{align}
&\Pi_1^{s_i}(p_i) \equiv \Pi^{ij}{}_{ik,jl,kl}(s_i,p_i) - \frac{1}{2} \Pi^{ij}{}_{ij,kl,kl}(s_i,p_i) \,, 
\hspace{2.5cm} \Pi^{s_i}(p_i) \equiv \Pi^{ii}{}_{lk,jl,jk}(s_i,p_i) \,, \nonumber \\
&\Pi_2^{s_i}(p_i) \equiv \Pi^{ij}{}_{jk,il,kl}(s_i,p_i) - \Pi^{ij}{}_{ij,kl,kl}(s_i,p_i) + \frac{p_1^2}{p_3^2} 
\Big[ 2\Pi^{ij}{}_{ik,jl,kl}(s_i,p_i) - \frac{1}{2} \Pi^{ij}{}_{ij,kl,kl}(s_i,p_i) \Big] \,. \label{Pi-I-def-app}
\end{align}
Note that after contracting all indices of \eqref{Pi-def-app}, first it becomes a function of the magnitude of the momenta only and second, it can be rewritten completely in terms of the ratios of the momenta defined in \eqref{p-ratio}.

Substituting \eqref{Pi-def-app} into \eqref{Pi-I-def-app}, and then using \eqref{phi-i}, after some manipulations, we find the following explicit expressions
\begin{align}
&\Pi_1^{s_1,s_2,s_3}(p_1,p_2,p_3) = -\frac{1}{64}
\Big[
-8 s_2 s_3 \sin\varphi_2+8 s_1 \sin (\varphi_2-\varphi_3) 
\left(s_2-s_3 \cos\left(\varphi_2-\varphi_3\right)\right)
 \\
&\hspace{6cm}
+\sin (2\varphi_2-3 \varphi_3)+3 \sin \left(2 \varphi_2-\varphi_3\right)+6 \sin \left(\varphi_3\right)
\Big] \sin\varphi_3 \,, \nonumber \\ 
&\Pi_2^{s_1,s_2,s_3}(p_1,p_2,p_3) = 
-\frac{1}{64 x_3^2} \Big[16 s_1 s_2 \sin(\varphi_2-\varphi_3)
-8 s_1 s_3 \sin 2(\varphi_2-\varphi_3)+8 s_1 s_2 x_3^2 \sin \left(\varphi _2-\varphi _3\right)
 \\
&\hspace{2cm}-4 s_1 s_3 x_3^2 \sin 2\left(\varphi_2-\varphi_3\right)- 4 s_2 s_3 
\left(1-x_3^2\right) \sin \left(\varphi_2-2 \varphi_3\right)
-12 s_2s_3 \left(1+x_3^2\right) \sin \left(\varphi_2\right)
\nonumber \\
&\hspace{2cm}
+3 \sin \left(2 \varphi_2-3 \varphi_3\right)+5 \sin \left(2 \varphi_2-\varphi_3\right)
+4 x_3^2 \sin\left(2 \varphi_2-\varphi_3\right)+6 \left(1+2 x_3^2\right) \sin\varphi_3 \Big] 
\sin\varphi_3 \,, \nonumber \\ 
&\Pi^{s_1,s_2,s_3}(p_1,p_2,p_3) = 
\frac{1}{32} \Big[4 \sin\varphi_3 \cos\varphi_2 \left(s_1 s_2 \sin\varphi_3+\sin\varphi_2
\left(\cos\varphi_3-s_1 s_3\right)\right)
 \\ \nonumber
&\hspace{1cm} +4 \sin\varphi_2 \left(s_1 s_3 \sin\varphi_2 \cos\varphi_3
+s_2 \sin\varphi_3 \left(s_3-s_1 \cos\varphi_3\right)\right)
+2 \cos 2\varphi_3 \cos^2\varphi_2
+\cos 2\varphi_2-3 \Big] \,.
\end{align}

In particular case of $s_1=s_2=s_3 = +$, they simplify to
\begin{align}
&\Pi_1^{+++}(x_2,x_3) = -\frac{\left(x_2-3 x_3-1\right) \left(x_2-x_3-1\right) \left(x_2-x_3+1\right)
\left(x_2+x_3-1\right){}^2 \left(x_2+x_3+1\right){}^3}{256 x_2^2 x_3^4}
\,, \nonumber \\ 
&\Pi_2^{+++}(x_2,x_3) = \frac{\left(x_2-x_3-1\right) \left(x_2-x_3+1\right) \left(x_2+x_3-1\right){}^2
\left(x_2+x_3+1\right){}^3 \left(4 x_3^3+5 x_3-3 x_2+3\right)}{256 x_2^2 x_3^6} , \nonumber \\ 
&\Pi^{+++}(x_2,x_3) = \frac{\left(\left(x_2-x_3\right){}^2-1\right) \left(x_2+x_3-1\right)
\left(x_2+x_3+1\right){}^3}{64 x_2^2 x_3^2} \,,
\end{align}
and also for the case of $s_1=s_2=+$ and $s_3 = -$, we have
\begin{align}
&\Pi_1^{++-}(x_2,x_3) = \Pi_1^{+++}(x_2,-x_3) \,, 
&\Pi_2^{++-}(x_2,x_3) = \Pi_2^{+++}(x_2,-x_3) \,, \nonumber \\
&\Pi^{++-}(x_2,x_3) = \Pi^{+++}(x_2,-x_3) \,.
\end{align}

We use the above explicit expressions to find shapes of the bi-spectra for the graviton NGs.

{}

\end{document}